\documentclass[12pt]{article} 
\usepackage{amsmath}
\usepackage[dvips]{epsfig}
\usepackage{float}

\def\ga		{\alpha}
\def\gb		{\beta}
\def\gc		{\gamma}
\def\gd		{\delta}
\def\gee	{\epsilon}

\def\go		{\omega}
\def\gr		{\rho}
\def\gs		{\sigma}
\def\gt		{\theta}
\def\ek		{\gee_{k}}

\def\capo	{\right.\\ \left.}
\def\dn		{\downarrow}
\def\up		{\uparrow}
\def\la		{\langle}
\def\ra		{\rangle}

\renewcommand{\[}{\left[}
\renewcommand{\]}{\right]}
\renewcommand{\(}{\left(}
\renewcommand{\)}{\right)}

\begin{document}

\pagestyle{empty}

\vspace*{10mm}

\begin{center} 
{\Large\bf Three-Body and One-Body Channels  }\\[4mm] 
{\Large\bf of the  Auger Core-Valence-Valence decay:}\\[4mm]
{\Large\bf Simplified Approach  }\\[17mm]
{\large Andrea Marini and Michele Cini}\\[3mm] 
\normalsize
{\it Istituto Nazionale di Fisica della Materia, Dipartimento di 
Fisica,}\\[.6mm] 
{\it Universita' di Roma Tor Vergata, Via della Ricerca Scientifica, 1-00133}
\\[.6mm]
{\it Roma, Italy}\\[7mm]
\end{center}
\vskip 17mm
\begin{quote}
\begin{center}{\bf Abstract\\[1mm]}\end{center}
We propose a computationally simple  model of
Auger and APECS line shapes from open-band solids. Part of the intensity 
comes from the decay of unscreened core-holes and is obtained by the
two-body Green's function $G^{(2)}_{\go}$, as in the case of filled 
bands. The rest of the intensity arises from screened core-holes and is 
derived using a  variational description of the relaxed ground state; this 
involves the two-holes-one-electron  propagator $G_{\go}$, which also 
contains  one-hole contributions.  
For many transition metals, the two-hole Green's
function $G^{(2)}_{\go}$ can be well described by the  Ladder
Approximation, but the {\em three}-body Green's function poses serious further
problems. To calculate $G_{\go}$, treating electrons and holes on 
equal footing, we propose a practical approach  to sum the series to all 
orders. 
We achieve that by formally rewriting the problem in terms of a
fictitious three-body interaction. Our method grants non-negative 
densities of states, explains the apparent {\em negative-U} 
behavior of the spectra of early transition metals and interpolates well
between weak and strong coupling, as we demonstrate by test model
calculations.
\\[3mm]
\end{quote}

\newpage

\pagestyle{plain}
\pagenumbering{arabic}

\section{Introduction}
\setcounter{equation}{0}
The Core-Valence-Valence (CVV) Auger Spectra of solids with closed 
valence bands are well described~\cite{reviews} in terms of the two-holes 
Green's function $G^{(2)}_{\go}$ by the Cini~\cite{Cini:1976,Cini:1977} 
and Sawatzky~\cite{Sawatzky:1977} (CS) model.

The CS model allows to understand the phenomenology involving band-like, 
atomic-like and intermediate situations in terms of the $U/W$ ratio of 
the on-site repulsion $U$ to the band width $W$. For low $U/W$, the 
line shape is close to the self-convolution on the local one-hole 
density of states;  with increasing $U/W$, the shape is distorted 
until, for a critical value of the ratio, two-hole resonances appear. 
They correspond to poles of  $G^{(2)}_{\go}$, as calculated from Anderson,
Hubbard or related models. Atomic-like peaks and band-like structures 
are often observed~\cite{reviews} in the same spectrum. 
For  high $U/W$ (atomic-like case) the Auger line shapes are so close to
the free-atom spectra that they  are labeled by LSJ terms and levels.
Detailed studies of noble transition metals, like Au~\cite{Au} and
Ag~\cite{Ag} led to a very good agreement between theory and experiment,
and also allowed the direct observation of off-site interaction
effects~\cite{offsite}. 
A similar success was achieved for covalently bonded solids~\cite{cov}.
In all cases  the diagrammatic expansion of $G^{(2)}_{\go}$ is just a
ladder of successive interactions between the holes without self-energy
terms or vertex corrections.

For open valence bands, the theory is much more complicated, and it 
is known from experiment that  no such narrow atomic-like peaks exist 
any more. 
However for almost completely filled bands, when the number of holes per 
quantum state  $n_{h}\ll 1$, remnants of the atomic multiplet structure are
still observed  and the closed-band theory can be
extrapolated~\cite{Cini:1979}. 
To a first approximation, one can assume that the valence electrons remain 
frozen during the core ionisation, and in the initial state of the Auger 
decay  the valence configuration is the same as in the ground state. The 
intra-band shake-up effects are $O(n_{h})$ and can be accounted for by 
convolving with an asymmetric Doniach-\v{S}unji\'{c}~\cite{ds} line shape.
In this way, the CVV spectra are still described in terms of  the 
two-hole Green's function $G^{(2)}_{\go}$, computed in the absence of the core 
hole. 
Using Galitzkii's {\it Low Density Approximation}~\cite{Galitzkii:1958}
(LDA) the dominant diagrams of the perturbation expansion of the
$G^{(2)}_{\go}$ are just the same ladder diagrams which provide the exact
solution for $n_{h}\rightarrow 0$. This gives a satisfactory
explanation of the Auger spectra for $n_{h}\approx 0.1$, which includes
interesting cases like Ni and Pd~\cite{Cini:89}, even in finely dispersed
form~\cite{Cini:90}.  
To extend the LDA to larger $n_{h}$, a self-consistent version of the 
Low Density Approximation that involves using dressed propagators in the
ladder series was naturally  suggested~\cite{Drchal:1984}. 
However an unexpected result cames from cluster studies~\cite{Verdozzi:86}
which clearly demonstrated that the ladder approximation with bare
propagators is superior and is a good  approximation to $G^{(2)}_{\go}$ for
a wide $n_{h}$ range. This is the {\em Bare Ladder Approximation} (BLA).
This approach has been useful  to interpret the line shape of
Graphite~\cite{graphite}.

The next turn came from experiments on early $3d$ transition metals, 
like Ti and Sc~\cite{expUneg}, that  could not be interpreted by 
the above theory. 
The maximum of the line shape was shifted by the interaction to lower 
binding energy, which is the contrary of what happens in closed band 
materials. Qualitatively the CS model could work if one admitted that
$U<0$, and such an explanation has actually been 
proposed~\cite{propUneg}. 
 
However, no other evidence of $U<0$ was found; rather, it became clear that 
for almost empty bands one must formulate a new theory which is no 
simple extrapolation of the closed-band approach. 
Sarma~\cite{Sarma:92} first suggested that the Auger line shape 
of Ti looks like some linear combination of the one-electron density of 
states and its convolution. Using this hint, and a general formulation of 
the Auger decay by Gunnarsson and Sch\"{o}nhammer~\cite{Gunn:1980}, a 
simple explanation of the apparent {\em negative-U} behavior was 
found~\cite{Cini:94}. In this theory, the Auger line shape has two main
contributions, that we call unrelaxed and relaxed, respectively. The 
unrelaxed contribution is obtained assuming that the Auger decay 
occurs while the conduction electrons are in their ground state 
$|\psi\ra$ in the absence of the core hole. The density of states which 
shows up in this contribution is obtained by $G^{(2)}_{\go}$, like in 
closed-band systems. However, this is only a part of the story, and in many 
ways the {\em easy} part, because the screening electronic cloud that
surrounds the core-hole can partecipate in the Auger decay. The 
ground state $|\phi\ra$ in the presence of the core hole also enters 
the description.
The relaxed contribution is computed with $|\phi\ra$ as 
the initial state of the Auger decay. We summarise in the Appendix
\ref{appendix1} the argument wich leads us to express the relaxed 
contribution in terms of a three-body  (the Auger 
holes and the screening electron) density of states $D$; this is 
obtained from the Fourier transform of the $t>0$ part of the Green's function
\begin{align}
G\(\ga_l\gb_l\gc_l,\gc'_l\gb'_l\ga'_l;t\)=\(-i\)^3
\la\psi|T\left\{a_{\ga_l}^{\dag}\(t\)a_{\gb_l}^{\dag}\(t\)a_{\gc_l}\(t\)
a_{\gc'_l}^{\dag}a_{\gb'_l}a_{\ga'_l}\right\}|\psi\ra.
\label{eq:g3}
\end{align}
Here, operators are in the Heisenberg picture and we introduce a special 
notation $\ga_{l},\gb_{l}\ldots$ to indicate the set of quantum numbers of
the {\em local} valence spin-orbitals belonging to the atom where the Auger
decay occurs (the {\em Auger site}, for short). 
Experimentally, one can single out the relaxed contribution of 
(\ref{eq:g3}) by properly fixing the photoelectron energy in an
Auger-Photoelectron Coincidence Spectroscopy (APECS) experiment,
~\cite{Haak:1978,Haak:1984,Sawatzky:book,Jensen}
where the Auger electron is detected in coincidence with 
the photoelectron reponsible of the core hole creation. 
Fixing the photoelectron energy, the Auger electron measured in coincidence
comes from the decay of a few dominant intermediate states in the presence of
the core hole~\cite{Gunnarsson:82}.

Since  (\ref{eq:g3}) is  hard to calculate and even much harder than 
of $G^{(2)}_{\go}$,  we propose  a simple approach in 
the spirit of the BLA. If we wish to design an affordable scheme in 
this difficult problem, we must be prepared to adopt a \underline{series}
of approximations: Cini model, Ladder Approximation and a new one, which
we call the Core Approximation. 
Section \ref{pe} is devoted to the fomulation of the new 
scheme. The degree of validity of our approach will be investigated by 
comparison with  exact results from cluster calculations in Section 
\ref{cluster}.

\section{Perturbation expansion of the two and three-body \\
Green's functions}\label{pe}
\setcounter{equation}{0}
In the Cini model, we express the Auger line shape in terms of 
{\em local} Green's functions, like the one of Equation (\ref{eq:g3});
moreover, we calculate them by considering 
only the {\em local} scattering at the Auger site, since in this way 
we drastically simplify the algebra, and the line shape is little 
influenced by scattering at the other sites~\cite{sl}. Many Auger line shape 
calculations in solids have been performed in this way. 

The local interaction approach has been extended 
successfully to open bands~\cite{Cini:89,Cini:90}. In 
the present work we want to study its application to more general 
$n_{h}$. This should allow to extend the analysis to several transition 
metals, giving at least a qualitative understanding of their 
spectra, which is currently a difficult task.

In order to properly evaluate the results one should bear in mind 
that currently even for  $G^{(2)}_{\go}$ we have reliable  recipes only for 
$n_{h}$ less than $\approx 0.25$. 
This problem involves one more body and highly excited 
states of interacting systems; consequently even the main features
of the solution are quite an unsettled question. 

\subsection{Ladder Approximation to the two-hole propagator}\label{th}
The time-dependent Green's function  $G^{(2)}_{t}$ allows the usual 
perturbation 
expansion, in terms of time-ordered products of 
interaction-representation operators, namely:
\begin{align}
G^{(2)}\(\ga_l\gb_l,\gb'_l\ga'_l;t\)=\sum_n\(-i\)^{n}\int_{-\infty}^{\infty}
dt_1\dots\int_{-\infty}^{\infty}dt_n
\la T\left\{a_{\ga_l}^{\dag}\(t\)a_{\gb_l}^{\dag}\(t\)
H_U\(t_1\)\dots H_U\(t_n\)a_{\gb'_l}a_{\ga'_l}\right\}\ra_c,
\label{expansion3}
\end{align}
with  the average taken over the non interacting ground state $|\psi_0\ra$ and
the sum restricted to the topologically inequivalent, connected 
diagrams. Ordering the spin-orbitals in an arbitrary way, we may 
write the Coloumb valence-valence interaction between those of the
Auger site in the form
\begin{align}
H_U=\sum_{\mu_{l}<\nu_{l},\gr_{l}
<\tau_{l}}U_{\mu_{l}\nu_{l}\gr_{l}\tau_{l}}a^{\dag}_{\mu_{l}}a^{\dag}_{\nu_{l}}
a_{\tau_{l}}a_{\gr_{l}}\label{eq:core0}.
\end{align}
The diagrammatic method develops Equation (\ref{expansion3}) in 
terms of local non-interacting time ordered one-body propagator 
$S_0\(\ga_l,\gb_l;t\)$ 
\begin{align}
S_0\(\ga_l,\gb_l;t\)=
S_0^h\(\ga_l,\gb_l;t\)-S_0^e\(\gb_l,\ga_l;-t\),
\label{s0}
\end{align}
where
\begin{gather}
S_0^h\(\ga_l,\gb_l;t\)=-i\gt\(t\)\la a^{\dag}_{\ga_l}\(t\)a_{\gb_l}\ra,\\
S_0^e\(\gb_l,\ga_l;-t\)=-i\gt\(-t\)\la a_{\gb_l}a^{\dag}_{\ga_l}\(t\)\ra.
\end{gather}
here the average is taken over the non-interacting ground state 
$|\psi_0\ra$ with energy $E_0$.
For $n_{h}<0.25$ and a range of $U/W$, a good 
start is provided by the BLA. In the BLA, one selects the series of ladder 
diagrams 
which are free of self-energy insertions and vertex corrections.
This is equivalent to the approximate factorisation
\begin{multline}
\la T\left\{a_{\ga_l}^{\dag}\(t\)a_{\gb_l}^{\dag}\(t\)
H_U\(t_1\)\dots H_U\(t_n\)a_{\gb'_l}a_{\ga'_l}\right\}\ra_c
\approx 
\sum_{\mu_{l}<\nu_{l},\gr_{l}<\tau_{l}}U_{\mu_{l}\nu_{l}\gr_{l}\tau_{l}}\\
\[\la T\left\{a_{\ga_l}^{\dag}\(t\)a_{\gb_l}^{\dag}\(t\)
a_{\tau_{l}}\(t_{1}\)a_{\gr_{l}}\(t_{1}\)\right\}\ra
\la T\left\{a_{\mu_l}^{\dag}\(t_{1}\)a_{\nu_l}^{\dag}\(t_{1}\)
H_U\(t_2\)\dots H_U\(t_n\)a_{\gb'_l}a_{\ga'_l}\right\}\ra_c\].
\label{bla1}
\end{multline}
Since
\begin{align}
g\(\ga_l\gb_l,\tau_l\gr_l;t-t_{1}\)=
\la T\left\{a_{\ga_l}^{\dag}\(t\)a_{\gb_l}^{\dag}\(t\)
a_{\tau_{l}}\(t_{1}\)a_{\gr_{l}}\(t_{1}\)\right\}\ra, 
\end{align}
is the non-interacting Green's function, the BLA leads to
\begin{multline}
G^{(2)}\(\ga_l\gb_l,\gb'_l\ga'_l;t\)=g\(\ga_l\gb_l,\gb'_l\ga'_l;t\)\\
(-i)\sum_{\mu_{l}<\nu_{l},\gr_{l}
<\tau_{l}}U_{\mu_{l}\nu_{l}\gr_{l}\tau_{l}}
\int_{-\infty}^{\infty}dt_{1}
g\(\ga_l\gb_l,\tau_l\gr_l;t-t_{1}\)G^{\(2\)}\(\mu_l\nu_l,\gb'_l\ga'_l;t_{1}\).
\label{bla2}
\end{multline}
Going to frequency space, this becomes a linear algebraic system.
Equation (\ref{bla2}) is called {\em Bare} Ladder Approximation because 
it uses undressed single particle propagators. It is the exact solution  
for $n_{h}=0$, and remains a good approximation for a useful range of 
$n_{h}$. It is simply equivalent to keeping only those diagrams that
remain in the closed-band limit. For closed bands, the one-body
propagators of Equation (\ref{s0}) reduce to the $S^{h}_{0}$ part;
therefore, lines that start as hole lines never go back in time 
and remain hole lines thoughout the diagrams. So, the order of times 
$t,t_{1},t_{2},\ldots,0$  remains fixed (decreasing) despite the
presence of the time-ordering operator T. 

The convolution form of Equation (\ref{bla2}) has furter important
consequences.
Mathematically, it is a Dyson equation in which the $U$ matrix is an
instantaneous self-energy. Therefore, it grants the Herglotz property:
for any interaction strenght, $G^{(2)}$ generates non-negative densities
of states. The Herglotz property is a basic requirement for a sensible
approximation, yet it is not easily obtained by diagrammatic approaches.
Its achievement is one of the most interesting features of this simple 
approximation.
 
\subsection{Ladder Approximation to the three-body propagator}\label{tb}
The Green's function  (\ref{eq:g3}) yields the expansion
\begin{multline}
G\(\ga_l\gb_l\gc_l,\gc'_l\gb'_l\ga'_l;t\)=
\sum_n\(-1\)^n\(-i\)^{3n+3}\int_{-\infty}^{\infty}
dt_1\dots\int_{-\infty}^{\infty}dt_n\\
\la T\left\{a_{\ga_l}^{\dag}\(t\)a_{\gb_l}^{\dag}\(t\)a_{\gc_l}\(t\)
H_U\(t_1\)\dots H_U\(t_n\)a_{\gc'_l}^{\dag}a_{\gb'_l}a_{\ga'_l}\right\}\ra_c.
\label{expansion}
\end{multline}
This describes the propagation of two-holes and one electron in the 
final state, or, if the electron and one hole annihilate, a one-body 
propagation results (see Section \ref{picco}). 
In proposing an approximation to Equation (\ref{expansion}), we may 
proceed by analogy with $G^{2}_{\go}$.  For $G_{\go}$ it is natural to 
propose a bare ladder approximation in which the lines that start as
electron (hole) lines never go back in time and remain electron (hole) lines
throughout the diagrams; such an approximation should have essentialy
the same physical contents as in the previous case and work properly in a 
wide range of $U$ and fillings. However, the series cannot be summed 
easily like in Equation (\ref{bla2}), 
because with three  bodies involved we meet an extra difficulty. For 
instance, let $H_U\(t_i\)$ produce an interaction between the two 
holes in a given term of the expansion; then, the electron line 
overtakes time $t_i$; therefore the diagram does {\em not} yield a 
convolution of a function of $t-t_i$ times a function of  $t_i$.
This undesirable feature can be removed by using the identities
\begin{align}
S_0^h\(\ga_l,\gb_l;t\)=
i\sum_{\gc}S_0^h\(\ga_l,\gc;t-t'\)S_0^h\(\gc,\gb_l;t'\).
\label{eq:core2}
\end{align}
and
\begin{align}
S_0^e\(\ga_l,\gb_l;t\)=
i\sum_{\gc}S_0^e\(\ga_l,\gc;t-t'\)S_0^e\(\gc,\gb_l;t'\).
\tag{\ref{eq:core2}$'$}
\end{align}
where the summations run over all the complete set of spin-orbitals. These 
identities are derived in Appendix \ref{appendix2}. Here we note that in the 
limit $t=t^{\prime}\rightarrow 0^{+}$, we get
\begin{align}
S_0^h\(\ga_l,\gb_l;t\)\rightarrow 
-i\sum_{\gc}\la a^{\dagger}_{\ga}a_{\gc}\ra\la a^{\dagger}_{\gc}a_{\gb}\ra
\label{eq:compl},
\end{align}
and
\begin{align}
S_0^e\(\ga_l,\gb_l;t\)\rightarrow
i\sum_{\gc}\la a_{\ga_{l}}a^{\dag}_{\gc}\ra\la a_{\gc}a^{\dag}_{\gb_{l}}\ra.
\tag{\ref{eq:compl}$'$}
\end{align}
Since the ground state is not the hole vacuum, 
$\la a_{\gc}a^{\dag}_{\gb}\ra \neq \gd(\gc ,\gb)$; so the correct limits are 
obtained only thanks to the completeness of the $\gc$ set. 

\begin{figure}[t]
\begin{center}
\epsfig{figure=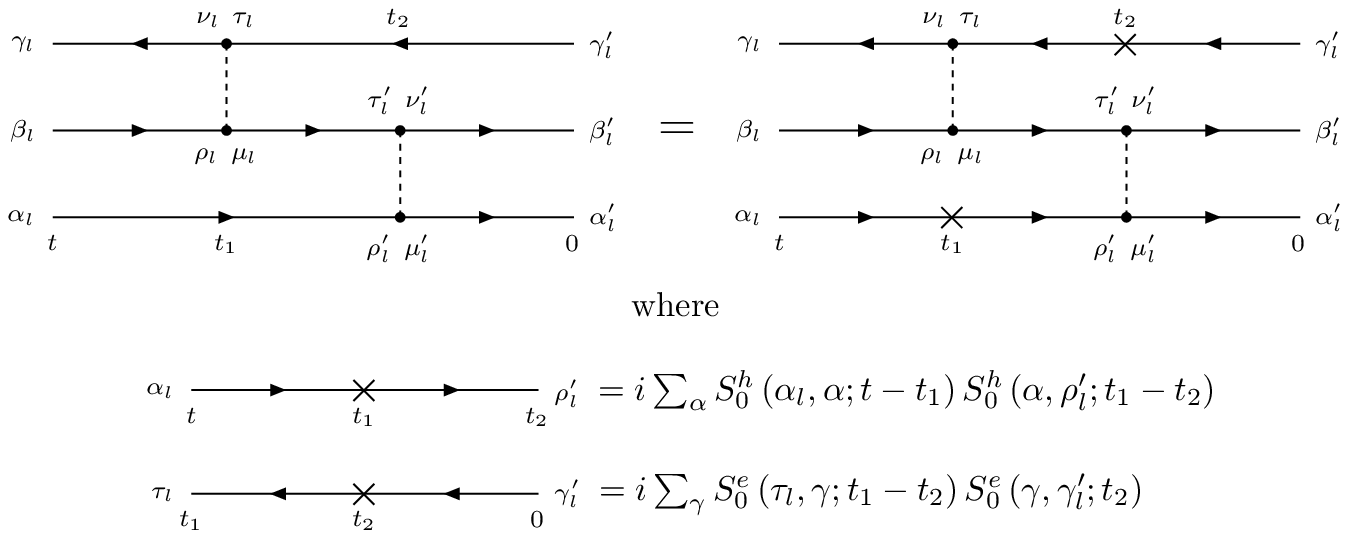,clip=,bbllx=75,bblly=632,bburx=479,bbury=802}\\
\caption{\footnotesize{
Second order contribution to the three-body Green's function.
Using Equations (\ref{eq:core2}) we cast it in the form of a
product of three ``blocks''. These are easly dealt with by a Fourier 
transform.}}\label{fig:1}
\end{center}\end{figure}

To see the use of Equations (\ref{eq:core2}), consider for instance the 
application to one of the second-order contributions
to (\ref{expansion}).
Using the standard diagrammatic rules, the l.h.s. of the uppermost Equation 
in Figure (\ref{fig:1}) reads (understanding a sum over repeated indices):
\begin{multline}
G_2\(\ga_l\gb_l\gc_l,\gc'_l\gb'_l\ga'_l;t\)=
-U_{\mu_{l}\nu_{l}\gr_{l}\tau_{l}}U_{\mu'_{l}\nu'_{l}\gr'_{l}\tau'_{l}}\(-i\)^2
\int\int_{-\infty}^{\infty}dt_1\,dt_2 \\
\[S_0^e\(\gc_l,\nu_l;t-t_1\)S_0^h\(\gb_l,\gr_l;t-t_1\)
S_0^h\(\ga_l,\gr'_l;t-t_2\)
\capo S_0^h\(\mu_l,\tau'_l;t_1-t_2\)S_0^e\(\tau_l,\gc'_l;t_2\)
S_0^h\(\nu'_l,\gb'_l;t_2\)S_0^h\(\mu'_l,\ga'_l;t_2\)\],
\label{eq:coreprimo3}
\end{multline}
Using the identities (\ref{eq:core2}), one casts Equation
(\ref{eq:coreprimo3})  in the convolution form 
\begin{multline}
G_2\(\ga_l\gb_l\gc_l,\gc'_l\gb'_l\ga'_l;t\)=
-U_{\mu_{l}\nu_{l}\gr_{l}\tau_{l}}U_{\mu'_{l}\nu'_{l}\gr'_{l}\tau'_{l}}\\
\sum_{\ga,\gc}\int\int_{-\infty}^{\infty}dt_1\,dt_2
\[G_0\(\ga_l\gb_l\gc_l,\nu_l\gr_l\ga;t-t_1\)
G_0\(\ga\mu_l\tau_l,\gc\tau'_l\gr'_l;t_1-t_2\)
G_0\(\mu'_l\nu'_l\gc,\gc'_l\gb'_l\ga'_l;t_2\)\].
\label{eq:core3}
\end{multline}
where 
\begin{align}
G_0\(\ga\gb\gc,\gc'\gb'\ga';t\)=
S_0^e\(\gc,\gc';t\)S_0^h\(\gb,\gb';t\)S_0^h\(\ga,\ga';t\).
\label{eq:core4}
\end{align}
This is the expression, that we represent pictorally as the  r.h.s. of 
the uppermost Equation in Figure (\ref{fig:1}).
In this way, introducing a fictitious $\times$ interaction vertex, along with 
the true interaction vertex (dot), we obtain a factorisation similar 
to Equation (\ref{bla1}) and the diagram is cast in the convolution 
form. This useful property extends to all the diagrams of the bare-ladder
approximation.

\subsection{Core-Approximation}\label{ca}
We achieved  Equation (\ref{eq:core3}) in the form of a convolution, which 
is a big simplification. However, the infinite summations (one for 
each $\times$ interaction) are a high price to pay for that. They arise
because  Equations (\ref{eq:core2}) imply a summation over the complete set
of $\gc$. On the other hand, since we use a {\em local} $H_{U}$,
the dot interaction involves only {\em local} matrix elements between 
spin-orbitals; so, we are only interested in the {\em local} 
elements $S^{h,e}\(\ga_{l},\gb_{l};t\)$. Physically, we may expect that 
only the sites which are closest to the Auger site give an important 
contribution to the summations, and we can actually work with a 
limited set to express the local $S^{h,e}\(\ga_{l},\gb_{l};t\)$ elements.
Larger sets will lead to more precise results, at the 
cost of more computation. Here,  we wish to explore the simplest 
approximation, by drastically limiting the $\gc$ summation to the 
local states $\gc_{l}$.
We have observed above that summing over the complete $\gc$ set is
necessary to get the correct zero-time limit. In turn, this is essential
to preserve normalisation.
If we simply replace $\gc$ with $\gc_l$  in Equations  (\ref{eq:core2}) 
this condition is violated (see Equations (\ref{eq:compl}));
therefore we introduce a set of functions 
$R^{\pm}\(\ga_l,\gb_l;t\)$ which approximately factor the propagator 
in analogy with  (\ref{eq:core2}) according to the ansatz
\begin{align}
(-i)\la a^{\dag}_{\ga_{l}}\(t\)a_{\gb_{l}}\ra\approx 
\sum_{\gc_{l}}R^{+}\(\ga_{l},\gc_{l};t-t'\)
\la a^{\dag}_{\gc_{l}}\(t'\)a_{\gb_{l}}\ra,
\label{eq:coreans}
\end{align}
and
\begin{align}
(-i)\la a_{\ga_{l}}\(t\)a^{\dag}_{\gb_{l}}\ra\approx
\sum_{\gc_{l}}R^{-}\(\ga_{l},\gc_{l};t-t'\)
\la a_{\gc_{l}}\(t'\)a^{\dag}_{\gb_{l}}\ra,
\tag{\ref{eq:coreans}$'$}
\end{align}
where $t'$ is any time intermediate between 0 and $t$;  the 
$R^{\(\pm\)}$ functions are computed for any t by solving the system
\begin{align}
(-i)\la a^{\dag}_{\ga_{l}}\(t\)a_{\gb_{l}}\ra=
\sum_{\gc_{l}}R^{+}\(\ga_{l},\gc_{l};t\)
\la a^{\dag}_{\gc_{l}}a_{\gb_{l}}\ra,
\label{eq:coreans2}
\end{align}
and
\begin{align}
(-i)\la a_{\ga_{l}}\(t\)a^{\dag}_{\gb_{l}}\ra=
\sum_{\gc_{l}}R^{-}\(\ga_{l},\gc_{l};t\)
\la a_{\gc_{l}}a^{\dag}_{\gb_{l}}\ra.
\tag{\ref{eq:coreans2}$'$}
\end{align}
The system must be identically satisfied for any t and in particular 
the correct $t\rightarrow 0^+$ limit is granted.

Equations (\ref{eq:coreans2}) are correct in the limit of core states, when 
$S^{h,e}\(\ga_{l},\gb_{l};t\)$ is diagonal in its indices and 
coincides with $R^{\pm}$; therefore we call this the 
{\em Core Approximation} (CA). 
The ansatz is also correct in the strong coupling case, when localised
two-hole resonances develop. This is appealing, since the strong coupling
case is the hard one, while at weak coupling practically every reasonable
approach yields similar results. 
Thus, we regard the ansatz (\ref{eq:coreans}) as a physically 
motivated aproximation, which must be tested against exact results 
for its validation. 

\subsection{Summing the Three-body ladder}\label{cla}
Working out the Core Approximation (CA) like in the example
(\ref{eq:core3}) one can compute all kinds of ladder diagrams, to all
orders. The partial sum of the series (\ref{expansion}) that one obtains in
this way will be referred to as  {\em Core-Ladder-Approximation} (CLA). 
From now on only {\em local} indices appear so we shall dispense 
ourselves from showing this explicitly.
In the exact expansion (\ref{expansion}), with the local $H_{U}$ denoting
the interaction Hamiltonian (\ref{eq:core0}) between fermions at the Auger
site, the outgoing holes at time t are labeled $\ga$ and $\gb$ (creation 
operators) and the outgoing electron $\gc$ (annihilation operator).  
In the CA for the n-th term of  (\ref{expansion}),
\begin{multline}
G_n\(\ga\gb\gc,\gc'\gb'\ga';t\)
=\(-1\)^n\(-i\)^{3n+3} \int_{-\infty}^{\infty}
dt_1\int_{-\infty}^{\infty}dt_{2}\dots\int_{-\infty}^{\infty}dt_n
\sum_{\mu<\nu,\gr<\tau}U_{\mu\nu\gr\tau}\\
\la T\left\{a_{\ga}^{\dag}\(t\)a_{\gb}^{\dag}\(t\)a_{\gc}\(t\)
a^{\dag}_{\mu}\(t_1\)
a^{\dag}_{\nu}\(t_1\)a_{\tau}\(t_1\)a_{\gr}\(t_1\)
H_U\(t_2\)\dots H_U\(t_{n}\)
a^{\dag}_{\gc'}a_{\gb'}a_{\ga'}\right\}\ra_c,
\label{eq:cla2}
\end{multline}
we consider the times ordered with $t\ge t_{1}\ge\ldots\ge 0$;
let us first analyse the contribution which arises when  $H_{U}(t_{1})$
describes hole-hole scattering.
In the expansion of the T-Product of Equation (\ref{eq:cla2})
by  Wick's theorem we  contract the $\ga$ and $\gb$ 
creation operators of the outgoing holes with the annihilation 
operators at time $t_{1}$, obtaining among other terms:
\begin{align}
\la T\left\{a_{\ga}^{\dag}\(t\)a_{\gr}\(t_1\)\right\}\ra
\la T\left\{a_{\gb}^{\dag}\(t\)a_{\tau}\(t_1\)\right\}\ra
\la T\left\{a_{\gc}\(t\)a^{\dag}_{\mu}\(t_1\)a^{\dag}_{\nu}\(t_1\)
H_U\(t_2\)\dots a^{\dag}_{\gc'}a_{\gb'}a_{\ga'}\right\}\ra_c,
\label{eq:cla3}
\end{align}
This, however cannot yet be factored in the convolution form because the
last term of Equation (\ref{eq:cla3}) depends on $t$.
In the ladder approximation, the annihilation operator $a_{\gc}\(t\)$
can be contracted with any creation operator at whatever time
$t_2,\dots,t_n$ or 0. Using the CA we can insert a $\times$ vertex
at time $t_1$ on the $\gc$  electronic line (like in Figure 
(\ref{fig:1})); namely
\begin{multline}
\la T\left\{a_{\ga}^{\dag}\(t\)a_{\gr}\(t_1\)\right\}\ra
\la T\left\{a_{\gb}^{\dag}\(t\)a_{\tau}\(t_1\)\right\}\ra
\la T\left\{a_{\gc}\(t\)a^{\dag}_{\mu}\(t_1\)a^{\dag}_{\nu}\(t_1\)
H_U\(t_2\)\dots a^{\dag}_{\gc'}a_{\gb'}a_{\ga'}\right\}\ra_c =\\
\la T\left\{a_{\ga}^{\dag}\(t\)a_{\gr}\(t_1\)\right\}\ra
\la T\left\{a_{\gb}^{\dag}\(t\)a_{\tau}\(t_1\)\right\}\ra
\sum_{\xi}iR^{-}\(\gc,\xi;t-t_1\)\\
\la T\left\{a_{\xi}\(t_1\)a^{\dag}_{\mu}\(t_1\)a^{\dag}_{\nu}\(t_1\)
H_U\(t_2\)\dots a^{\dag}_{\gc'}a_{\gb'}a_{\ga'}\right\}\ra_c.
\label{eq:cla4}
\end{multline}
This result can be easily verified by expanding with Wick's theorem, 
and then applying Equations (\ref{eq:core2})  although the proof 
is somewhat lengthy. In Equation (\ref{eq:cla4}), the propagation 
between times $t$ and $t_{1}$ is described by the factor
\begin{align}
G_0\(\ga\gb\underline{\gc},\underline{\xi}\tau\gr;t-t_1\)=R^-\(\gc,\xi;t-t_1\)
S^h_0\(\gb,\tau;t-t_1\)S^h_0\(\ga,\gr;t-t_1\),
\label{eq:cla6}
\end{align}
this is similar to a non-interacting three-body propagator, except 
that the $S^e$ has been replaced by a $R^{-}$; as a shorthand notation we 
underline the electron indices that correspond to a $R^-$ factor. 
In a similar way $H_{U}(t_{1})$ can give other possible ladder
contributions to Equation (\ref{eq:cla2});
these come from  the interaction of the $\gc$ electron with one of the 
two holes; in this case
is the other hole that goes to a $\times$ vertex and we meet the factors
\begin{gather}
G_0\(\ga\underline{\gb}\gc,\gc'\underline{\gb'}\ga';t-t_1\)=
S_0^e\(\gc,\gc';t-t_1\)
R^+\(\gb,\gb';t-t_1\)S^h_0\(\ga,\ga';t-t_1\),\\
G_0\(\underline{\ga}\gb\gc,\gc'\gb'\underline{\ga'};t-t_1\)=
S_0^e\(\gc,\gc';t-t_1\)
S_0^h\(\gb,\gb';t-t_1\)R^+\(\ga,\ga';t-t_1\);
\label{eq:cla6a}
\end{gather}
where underline hole indices correspond to a $R^+$ contraction.

\begin{figure}[t]\begin{center}
\epsfig{figure=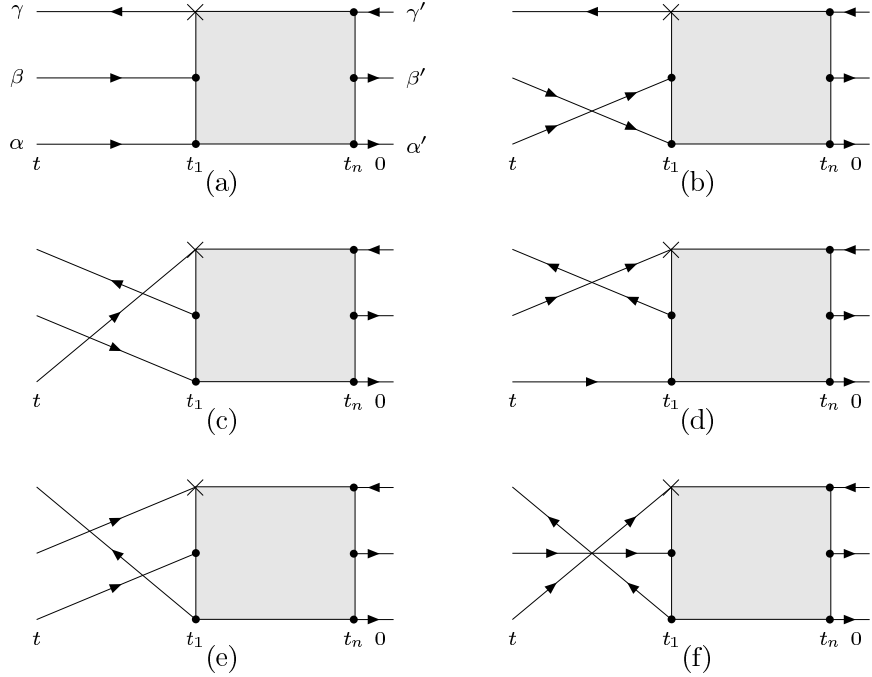,clip=,bbllx=118,bblly=493,bburx=451,bbury=762}\\
\caption{\footnotesize{
Diagrammatic representation of the contributions of order $n$ to the 
CLA (Equation (\ref{eq:cla7})).
The fictitious $\times$ interaction vertex represents a $R^{\pm}$ function
as explained in the example (\ref{eq:core3}).}}\label{fig:2}
\end{center}\end{figure}

A glance to Equation (\ref{eq:cla2}) reveals that the last factor in 
the rhs of Equation (\ref{eq:cla4}) is the 
matrix element which enters $G_{n-1}$; hence we obtain the CLA form for one of
the term of order $n$ of the development, corresponding to the diagram $(a)$
of Figure (\ref{fig:2})
\begin{align}
-\sum_{\mu<\nu,\gr<\tau}U_{\mu\nu\gr\tau}
\[\sum_{\xi} \int_{-\infty}^{\infty}d\,t_1\,
G_0\(\ga\gb\underline{\gc},\underline{\xi}\tau\gr;t-t_1\)
G_{n-1}\(\mu\nu\xi,\gc'\gb'\ga';t_1\)\].
\label{eq:cla5}
\end{align}

In other terms of the expansion,  $H_{U}(t_{1})$ describes the interaction
of the $\gc$ electron with the $\ga$ and $\gb$ holes; in addition, 
each contribution has an exchange counterpart, obtained by crossing the 
hole lines going from $t$ to $t_1$. 
Thus, applying the CA to the other contractions of eq.(\ref{eq:cla2})
we get the six contributions represented in the fig.(\ref{fig:2}).
Letting $n\rightarrow\infty$ and Fourier transforming we obtain a 
linear system of equations for the CLA form 
$G^{CLA}\(\ga\gb\gc,\gc'\gb'\ga';\go\)$ of the three-body Green's function, 
namely
\begin{multline}
G^{CLA}\(\ga\gb\gc,\gc'\gb'\ga';\go\)
=B_0\(\ga\gb\gc,\gc'\gb'\ga';\go\)-\\
\sum_{\xi}\left\{
\sum_{\mu<\nu}\[\sum_{\gr<\tau}U_{\mu\nu\gr\tau}
B_0\(\ga\gb\underline{\gc},\underline{\xi}\tau\gr;\go\)\]
G^{CLA}\(\mu\nu\xi,\gc'\gb'\ga';\go\)- \capo 
\sum_{\mu<\nu,\gr<\tau}U_{\mu\nu\gr\tau}\[
B_0\(\ga\underline{\gb}\gc,\nu\underline{\xi}\gr;\go\)
G^{CLA}\(\mu\xi\tau,\gc'\gb'\ga';\go\)+
\right.\capo\left. 
B_0\(\underline{\ga}\gb\gc,\mu\tau\underline{\xi};\go\)
G^{CLA}\(\xi\nu\gr,\gc'\gb'\ga';\go\)\]\right\}.
\label{eq:cla7}
\end{multline}
where the $B_0$ functions are 
\begin{gather}
B_0\(\ga\gb\gc,\gc'\gb'\ga';\go\)=
G_0\(\ga\gb\gc,\gc'\gb'\ga';\go\)-G_0\(\gb\ga\gc,\gc'\gb'\ga';\go\),\notag \\
B_0\(\ga\gb\underline{\gc},\underline{\xi}\tau\gr;\go\)=
G_0\(\ga\gb\underline{\gc},\underline{\xi}\tau\gr;\go\)-
G_0\(\ga\gb\underline{\gc},\underline{\xi}\gr\tau;\go\),\label{eq:cla7a}\\
B_0\(\ga\underline{\gb}\gc,\nu\underline{\xi}\gr;\go\)=
G_0\(\ga\underline{\gb}\gc,\nu\underline{\xi}\gr;\go\)-
G_0\(\underline{\ga}\gb\gc,\nu\gr\underline{\xi};\go\),\notag\\
B_0\(\underline{\ga}\gb\gc,\mu\tau\underline{\xi};\go\)=
G_0\(\underline{\ga}\gb\gc,\mu\tau\underline{\xi};\go\)-
G_0\(\ga\underline{\gb}\gc,\mu\underline{\xi}\tau;\go\).\notag
\end{gather}

The second, third and fourth lines of (\ref{eq:cla7}) come respectively 
from the (a,b), (c,d) and (e,f) diagrams of Figure (\ref{fig:2});
while the first two contributions come from the hole-hole interaction the 
others come from electron-hole interactions and convey information on 
the screening effects due to the electronic cloud which forms as a 
response to the deep electron ionization.

\subsection{Single-particle contribution }\label{picco}
At the same level of approximation we must consider the case when 
one (or both) the holes produced by the Auger transitions has the 
same spin as the screening electron.
Consider the spin-diagonal components
$G\(\ga\,\gb\,\gc,\,\gc'\,\gb'\,\ga';t\)$, (with $\gs_{\ga'}=\gs_{\ga}$ and 
so on); when the hole $\gb$ has the same z spin
component as the  electron, then, contracting  $a^{\dag}_{\gb}\(t\)$ with 
$a_{\gc}\(t\)$ and $a_{\gb'}$ with $a^{\dag}_{\gc'}$ in (\ref{eq:cla2})
one obtains the extra contribution 
\begin{align}
G^{sp}\(\ga\gb\gc,\gc'\gb'\ga';t\)= \(-1\)\la a^{\dag}_{\gb}a_{\gc}\ra
\la a^{\dag}_{\gc'}a_{\gb'}\ra S\(\ga,\ga';t\).
\label{eq:picco2}
\end{align}
where $ S\(\ga,\ga';t\)$ stands for the time-ordered dressed 
one-body Green's function that can be expanded with
\begin{align}
S\(\ga,\ga';t\)=
\sum_n\(-i\)^n \int_{-\infty}^{\infty}
dt_1\dots\int_{-\infty}^{\infty}dt_n
\la T\left\{a_{\ga}^{\dag}\(t\)H_U\(t_1\)\dots H_U\(t_n\)
a_{\ga'}\right\}\ra _c.
\label{expansion2}
\end{align}
and summed with the Dyson's equation~\cite{Mattuck} in terms of proper
Self-Energy diagrams.

\begin{figure}[t]
\begin{center}
\epsfig{figure=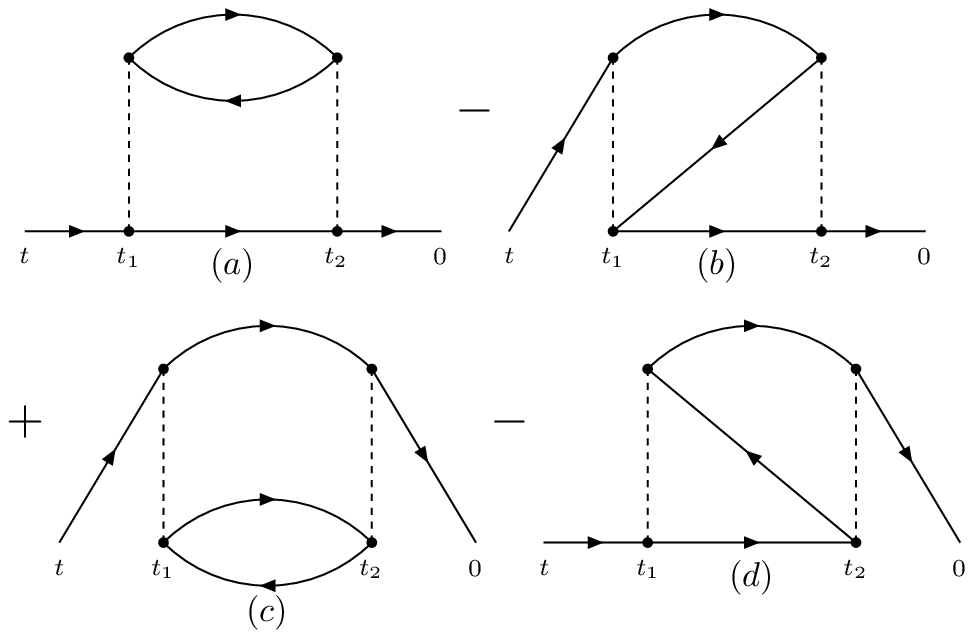,clip=,bbllx=154,bblly=577,bburx=447,bbury=774}
\caption{\footnotesize{
Topologically distinct contributions to the Second order self-energy.
No tadpole diagrams occur because we expand in terms of Hartree-Fock 
spin-orbitals.}}\label{fig:4}
\end{center}\end{figure}

At second order of (\ref{expansion2}) we find eigth diagrams
free of Hartree-Fock (tadpole) insertions that give rise to four topologically
distinct ones as shown in Figure  (\ref{fig:4}), which shows how 
the second-order self-energy can be
expressed in terms of the hole-hole-electron Green's function $G$. The rules
are:  (i) write down a diagram for G; (ii) join the incoming electron
line with one of the hole lines into an interaction vertex and draw an
interaction line to the other hole; (iii) repeat the operation with the 
outgoing lines; (iv) do all that in all possible ways.
This is a general result that can be extended to all orders yielding
the Dyson form for the $S\(\ga,\ga';\go\)$
\begin{multline}
S\(\ga,\ga';\go\)=S_0\(\ga,\ga';\go\)- \sum_{\mu'<\nu',\gr'<\tau'}
\sum_{\mu<\nu,\gr<\tau}U_{\mu\nu\gr\tau}U_{\mu'\nu'\gr'\tau'}\\
\left\{\[S_0\(\ga,\gr;\go\)G\(\mu\nu\tau,\nu'\tau'\gr';\go\)-
S_0\(\ga,\tau;\go\)G\(\mu\nu\gr,\nu'\tau'\gr';\go\)\]
S\(\mu',\ga';\go\)+\capo
\[S_0\(\ga,\tau;\go\)G\(\mu\nu\gr,\mu'\tau'\gr';\go\)-
S_0\(\ga,\gr;\go\)G\(\mu\nu\tau,\mu'\tau'\gr';\go\)\]
S\(\nu',\ga';\go\)\right\};
\label{eq:picco3}
\end{multline}
at the second order this reduces to the four terms of Figure (\ref{fig:4}).
We get a conserving approximation for the proper 
self-energy, using the CLA approximation for the three-body Green's 
functions of Equation (\ref{eq:picco3}).

By solving Dyson's equation (\ref{eq:picco3}) 
one can model XPS spectra from valence bands with low band filling; 
this is another field of application of our approach.

Having computed $S$, one finds how the Fermi energy $E_{F}$ is renormalised 
by the interaction. In general, we shall reference all the three-body 
spectra to the value of $E_{F}$ which are obtained by the 
corresponding one-body calculation. In the next Section, when comparing the 
line shapes with exact diagonalisation results, we shall align the Fermi 
levels accordingly.

The deep hole attracts a screening electron that can be directly 
involved in the Auger decay; this is the physical origin of the contribution
(\ref{eq:picco2}) to the three-body Green's function. Locally, 
such processes leave the system with  one hole in the final state.
The presence of an one-body contribution in the Auger 
spectra from transition metals like Ti or Sc has been pointed 
out already ~\cite{Sarma:92,Cini:94}. 
Besides the three-body (\ref{eq:cla7}) and one-body (\ref{eq:picco2}) 
diagrams, there are mixed contributions. Physically they represent 
interference contributions in which the system evolves from one-hole 
states to two-hole one electron states and back. Here we are going to 
neglect those diagrams for the sake of simplicity. Such terms can 
only be significant if the interaction is strong enough; however for 
strong interaction the one and three-body contributions are widely 
separated in frequency, so we may argue that the interference terms are 
not very important in general. Thus, we write
\begin{align}
G\(\ga\gb\gc,\gc'\gb'\ga';\go\)=G^{CLA}\(\ga\gb\gc,\gc'\gb'\ga';\go\)+
G^{sp}\(\ga\gb\gc,\gc'\gb'\ga';\go\).
\label{eq:picco5}
\end{align}

\section{Comparison with exact diagonalisation results}\label{cluster}
\setcounter{equation}{0}
In this Section, we wish to test the CLA results against those of a 
model system that can be diagonalised exactly.

\subsection{Model Cluster}
We consider a 5 atom cluster, with 2 levels for each atom; the one-body
basis elements are  $|si\gs\ra$ with $s=1,\dots,5$ the site index, $i=1,2$ the
level index and $\gs=\up,\dn$ for the spin direction; the one-body energies
are denoted by $\gee_{sj}$. 
The atoms $1\ldots4$ occupy the vertices of a square, and the Auger atom 
is at site 5 above the centre. Although this is just a model, we wish 
to use it to outline a possible procedure for the analysis of actual 
experimental data. Suppose we know a simple one-body Hamiltonian 
$H_{0}$ of the system, of the same level of sophistication as a tight-binding 
model of a solid, with a nearest 
neighbor hopping term:
\begin{align}
H_0=\sum_{si\gs}n_{si\gs}\gee_{si}+
\sum_{ij \gs}\sum_{\la ss'\ra} T^{ss'}_{ij} 
a_{si\gs}^{\dag}a_{s'j\gs},\label{eq:cluster3}
\end{align}
with real $ T^{ss'}_{ij}$. For nearest neighbor sites we assume 
identical matrices
\begin{align}
\left\{{\bf  T}_{ij}\right\}=
\begin{pmatrix}
t_{11} & t_{12} \\ t_{12} & t_{11}
\end{pmatrix},
\end{align}
Next, suppose we have an estimate of the Hubbard $U$ and we wish to 
model  the interactions by 
\begin{gather}
H_{Hub}=U \[\sum_{sij} n_{si\up}n_{sj\dn}+
\sum_{s\gs}n_{s1\gs}n_{s2\gs}\label{eq:cluster4}\].
\end{gather}
The full interacting model Hamiltonian cannot be taken to be the sum 
of $H_0$ and $H_{Hub}$, because $H_0$ must contain the effects of the 
interactions on the one-body states, roughly, at the Hartree-Fock 
level. A mean field average of the interaction must be subtracted out. 
Therefore, we assume the model Hamiltonian 
\begin{equation}
H=H_0+H_1,
\label{eq:cluster2}
\end{equation}
where
\begin{align}
H_1=H_{Hub}-V_{H-F}, 
\label{eq:ricluster4}
\end{align}
and
\begin{multline}
V_{H-F}=U\left\{\sum_{sij}\[\langle n_{si\up}\rangle n_{sj\dn}+
n_{si\up} \langle n_{sj\dn}\rangle\]+
\sum_{s\gs}\[\langle n_{s1\gs}\rangle n_{s2\gs}+
n_{s1\gs} \langle n_{s2\gs}\rangle + \right. \capo \left.
\(a^{\dag}_{s1\gs}a_{s2\gs}\la 
a_{s1\gs}a^{\dag}_{s2\gs}\ra+h.c.\)\]\right\}.
\label{eq:cluster5}
\end{multline}
where mean values are taken over the non-interacting ground state. 
In principle, if $H_{0}$ were a Self-Consistent-Field Hamiltonian, a 
self-consistent calculation of $\gee$ parameters and mean occupation 
numbers would be a more precise procedure. In this way, the Hartree-Fock
contribution to the self-energy 
would be automatically  enbodied in the bare propagators. In the 
present paper, we use non-interacting ground state averages, which is a cruder
approximation, because we wish to privilege the simplicity of the 
procedure;  the simplified approach 
already gives gratifying results (see below).

\subsection{Exact diagonalisation calculations}
We have chosen the following parameters in eV: $\gee_{s1}=0$ and 
$\gee_{s2}=1$ for all sites; $t_{11}=1.4$ and $t_{12}=0.4$.  In this way 
the bandwidth $W$, that is the difference between the extreme 
eigenvalues of $H_{0}$, is $\approx 9.6$ eV, a reasonable value for a 
transition metal. We want the densities $D$ which are related to the
positive time part of the Green's functions (\ref{eq:g3}), 
and we are particularly interested in the diagonal elements 
$D\(5i\gs_i,5j\gs_j,5k\gs_k;\go\)$; these are usually the largest and 
must be non-negative.

These are given by (dropping the site index $s=5$)
\begin{align}
D\(i\gs_i,j\gs_j,k\gs_k;\go\)=
\sum_n \left|\la\psi|a_{i\gs_i}^{\dag}a_{j\gs_j}^{\dag}a_{k\gs_k}
|E^{\(n\)}_{N-1}\ra\right|^2\gd\(\go+E^{\(n\)}_{N-1}-E\),
\label{eq:cluster6}
\end{align} 
where  N is the total number of electrons in the cluster,
$|\psi\ra$ is the interacting ground state with energy E; 
$|E^{\(n\)}_{N-1}\ra$ is  the complete set of interacting eigenstates
with $\(N-1\)$ electrons.
In one set of calculations we start from a ground state with $N=18$ valence 
electrons, since the maximum occupation in the cluster is 20, we have two
holes. 
We assume a non-magnetic ground state, with one hole for each spin; 
the configurations are  $\binom{10}{2}^2=100$. In the final 
state, with two Auger holes of opposite spin and a spin-up electron
we find  $\binom{10}{2}\binom{10}{1}=450$ intermediate states
with $\(N-1\)$ electrons.

In another set of calculations we start from a ground state with $N=16$,
that is, two holes for each spin, and get $|\psi\ra$ and $E$ by 
diagonalising a $\binom{10}{2}^2\times\binom{10}{2}^2=2025\times 2025$ 
matrix, while for $|E^{n}_{N-1}\ra$ we meet the maximum size of matrices
in our calculations, namely $\binom{10}{2}\binom{10}{3}=5400$.

The site $s=5$ was chosen to be the Auger atom because in this way 
the problem is highly symmetric, and we can classify all the states 
according to the Irreducible Representations $(IR)$ of the $C_{4v}$ Group, 
the Group of the square.
Carring out in Equation (\ref{eq:cluster6}) the summation on the  $IR$ of
the group we obtains 
\begin{align}
D\(i\gs_i,j\gs_j,k\gs_k;\go\)=
\sum_{IR}\sum_n \left|\la\psi|
a_{i\gs_i}^{\dag}a_{j\gs_j}^{\dag}a_{k\gs_k}|E^{\(IR,n\)}_{N-1}\ra\right|^2
\gd\(\go+E^{\(IR,n\)}_{N-1}-E\),
\label{eq:cluster7}
\end{align}  
but the operators $a_{i\gs}$ belong to the totalsymmetric representation 
$A_{1}$ of  and so $|\psi\ra$ and  $|E^{n}_{N-1}\ra$ must belong to the
same IR.
Since the ground state $|\psi\ra$ for $N=16$ and $N=18$ is  total-symmetric
we could restrict the summation to the  $A_{1}$ sector.
\begin{align}
D\(i\gs_i,j\gs_j,k\gs_k;\go\)=
\sum_n \left|\la\psi|a_{i\gs_i}^{\dag}a_{j\gs_j}^{\dag}a_{k\gs_k}
|E^{\(A_1,n\)}_{N-1}\ra\right|^2
\gd\(\go+E^{\(A_1,n\)}_{N-1}-E\).
\label{eq:cluster9}
\end{align}  
with a strong reduction of computing time.

\subsection{Comparison and evaluation of the CLA}
By the cluster calculations we want to study the physical consequences of 
the hole-electron interaction in the three-body density of states. 
Also, we wish to estimate the degree of validity of
the CLA.  We performed the  comparison with the exact results for several values
of $U/W$. 
By considering various possible electron populations $N\le 20$ in the
cluster we also varied the mean occupation of the Auger site in the
non-interacting ground state
\begin{align}
\langle n \rangle=\frac{1}{4}\sum_{\gs}\langle n_{51\gs}+n_{52\gs}\rangle.
\end{align}
The test becomes more severe when $\la n\ra$ is reduced towards 
half filling and $U/W$ is increased.
Many different densities of states are obtained from the matrix 
elements of $G(\go)$; in the context of our theory, the density
\begin{align}
D_{1h}(\go)\equiv
\la\psi|a_{51\up}^{\dag}a_{52\dn}^{\dag}a_{51\up}\gd\(\go+H-E\) 	
a_{51\up}^{\dag}a_{52\dn}a_{51\up}|\psi\ra.\label{eq:modello1}
\end{align}
is of special interest because in the diagrammatic series the annichilation
of  the spin-up electron  by the  hole of the same spin is particularly
strong so we may expect that the one-body term is important. 
By contrast, in the density 
\begin{align}
D_{2h1e}(\go)\equiv 
\la\psi|a_{51\up}^{\dag}a_{51\dn}^{\dag}a_{52\up}\gd\(\go+H-E\) 	
a_{52\up}^{\dag}a_{51\dn}a_{51\up}|\psi\ra.\label{eq:modello2}
\end{align}
the same contribution should be smaller and possibly absent.
We shall consider these two examples in turn. In Appendix \ref{appendixc} 
we detail the application of the CLA to the problem at hand. Although 
the analytic development can be somewhat boring, the maximum size of 
the matrices involved is just 8.

\begin{figure}[t]
\begin{center}
\epsfig{figure=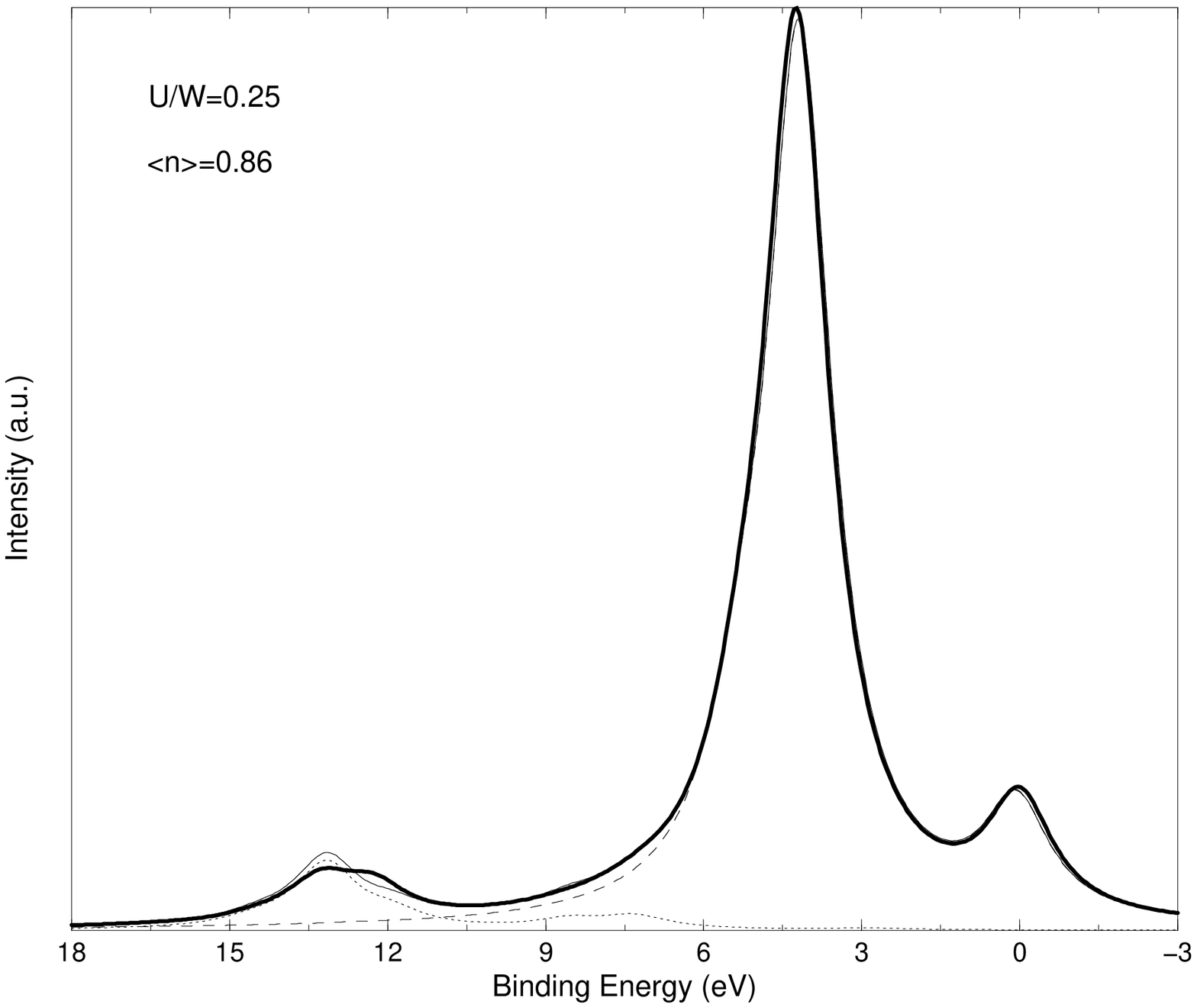,height=8cm,angle=-90}  
\epsfig{figure=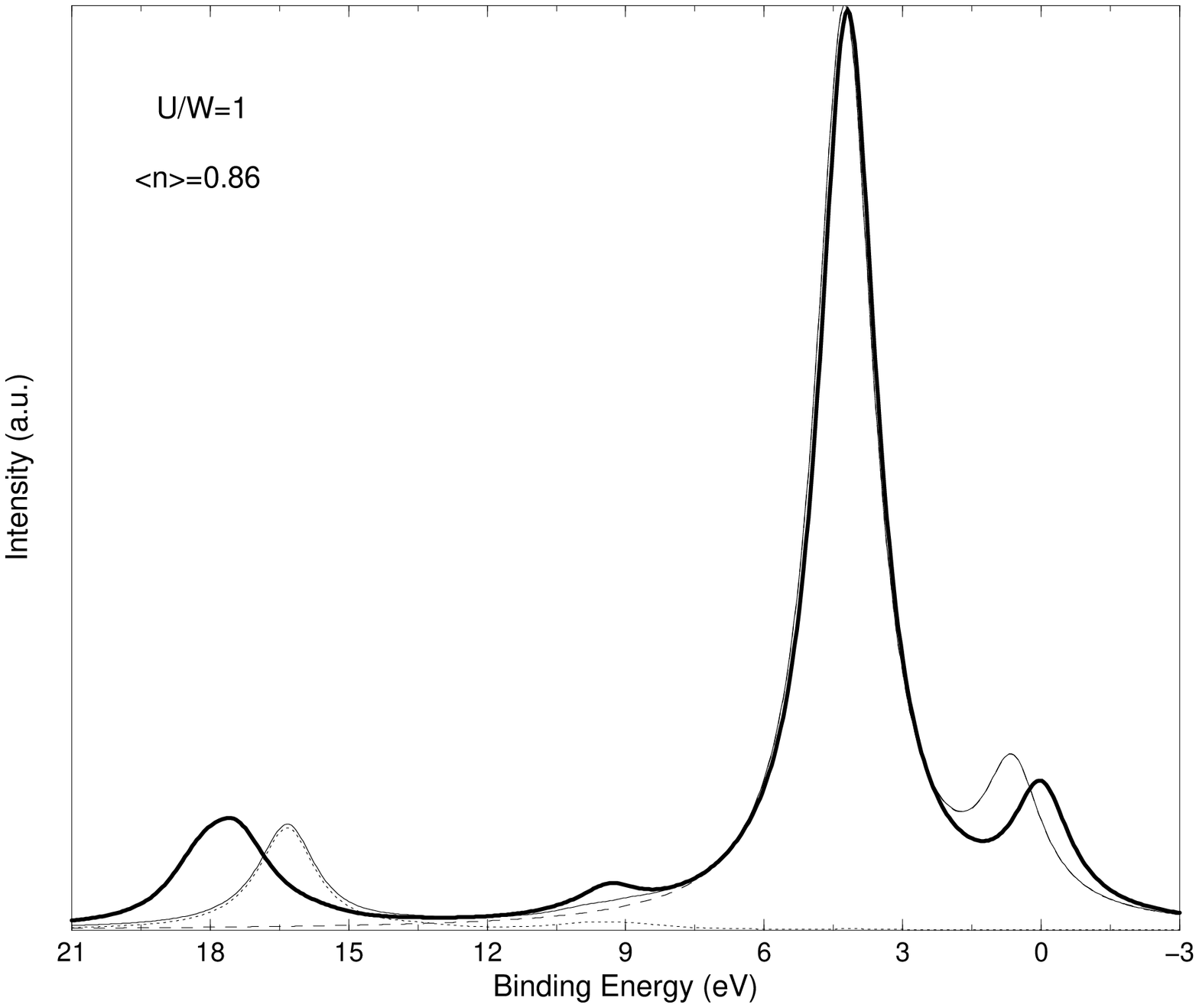,height=8cm,angle=-90}\\
\caption{\footnotesize{Binding energy dependence of $D_{1h}$ for
$\langle n\rangle=0.86$ with $U/W$=0.25 (left frame) and $U/W$=1 (right frame).
Heavy line: exact result; light: CLA;
dashed: 1-body contribution to the CLA result; dotted: 3-body
contribution to the CLA result. The line shapes have been 
convolved with a Lorentzian (FWHM=0.75 eV).}}\label{fig:5}
\end{center}\end{figure}

The left frame of Figure (\ref{fig:5}) shows $D_{1h}$ for $U/W$=0.25 
with $N=18$, which yields a population  $\la n\ra=0.86$ on the Auger site. 
The density is dominated by a single peak at binding energy $\approx 4$eV, 
but also shows a pair of wings. Since correlation effects are moderate and
the filling is fairly high, the CLA is still in good agreement with  the exact 
results.

In the right frame of Figure (\ref{fig:5}) the interaction is increased to
$U/W=1$. The pattern does not show any major changes, except a shift 
and an increase of the structure at high binding energy. The shift
($\approx 5$ eV) is large, which is understandable because the filling is
high and the screening uneffective.
The value of U is well outside the scope of weak coupling approaches
but the CLA still reproduces the exact results rather well.

Both the line shapes and the accuracy of the CLA are sensitive to
$\la n\ra$. In a series of calculations, we set $N=16$, which yields 
$\la n\ra=0.72$. 

\begin{figure}[H]
\begin{center}
\epsfig{figure=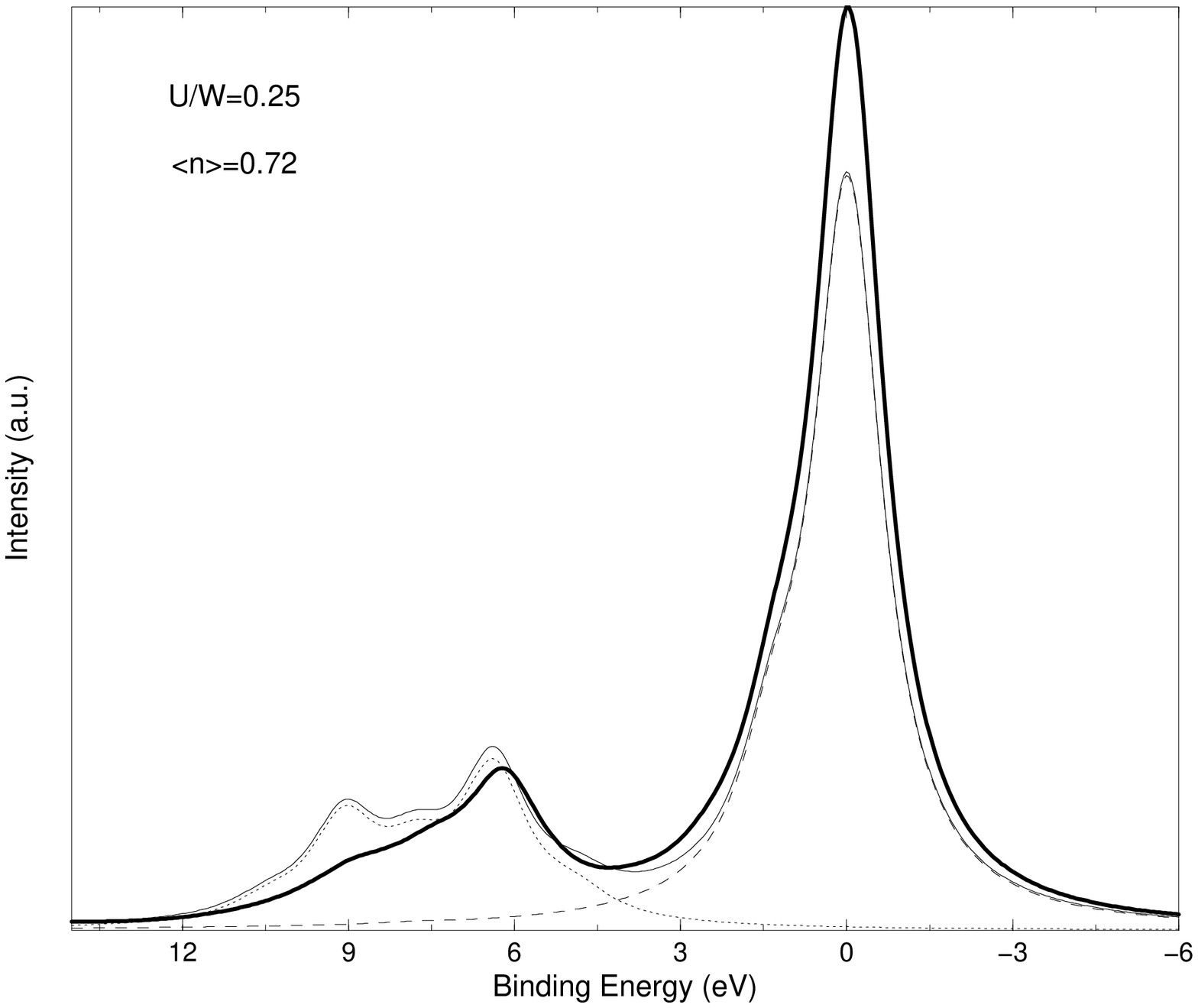,height=8cm,angle=-90}
\epsfig{figure=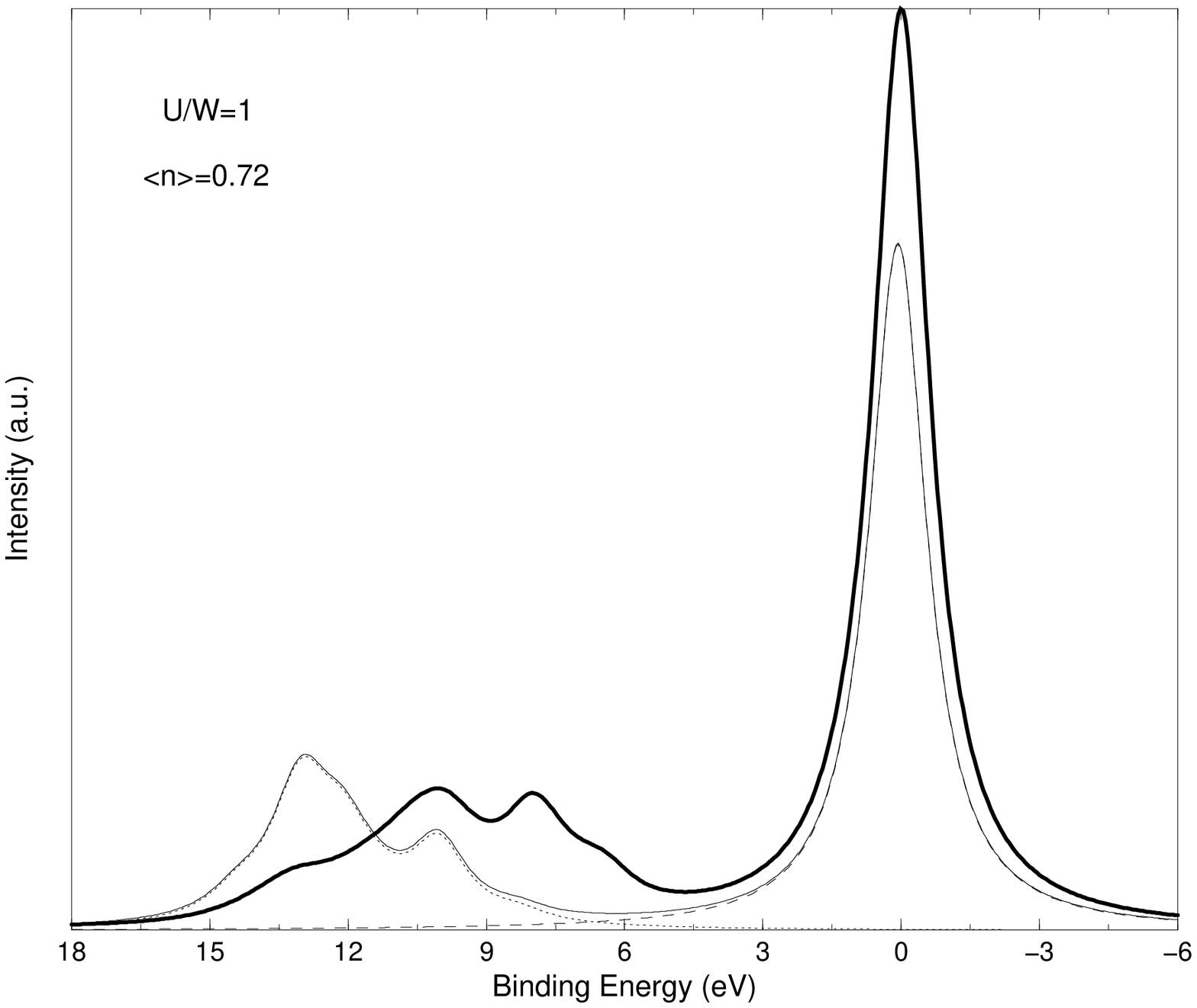,height=8cm,angle=-90}\\
\caption{\footnotesize{Binding energy dependence of $D_{1h}$ for  
$\langle n\rangle=0.72$ with $U/W$=0.25 (left frame) and $U/W$=1 (right frame).
Heavy line: exact result; light: CLA;
dashed: 1-body contribution to the CLA result; dotted: 3-body
contribution to the CLA result. The line shapes have been
convolved with a Lorentzian (FWHM=0.75 eV).}}\label{fig:6}  
\end{center}\end{figure}

In the left frame of Figure (\ref{fig:6}) a moderate interaction $U/W=0.25$ is 
assumed, and the exact results show a threshold peak centered at the
Fermi level accompanied by a rich structure 
between 5 and 11 eV. The CLA  underestimates somewhat the intensity of 
the main peak and does not faithfully reproduce the shape of the high 
binding energy structure, however the position of the main peak is 
correct and the overall line shape is in fair agreement. Moreover, 
the performance of the CLA does not break down quickly with 
increasing $U$ as weak-coupling approaches tend to do, but remains 
fairly stable. 
This can be seen in the right frame of Figure (\ref{fig:6}), 
where the comparison is done with $U/W$=1. 
Due to the effective screening in this case, the increase of $U$ does 
not cause a dramatic shift of the high binding-energy structure; this 
is borne out by the CLA and we may 
still speak of semiquantitative agreement. The CLA also explains the 
increase of the relative weight of the 3-body contribution with 
reducing band filling. 

\begin{figure}[t]
\begin{center}
\epsfig{figure=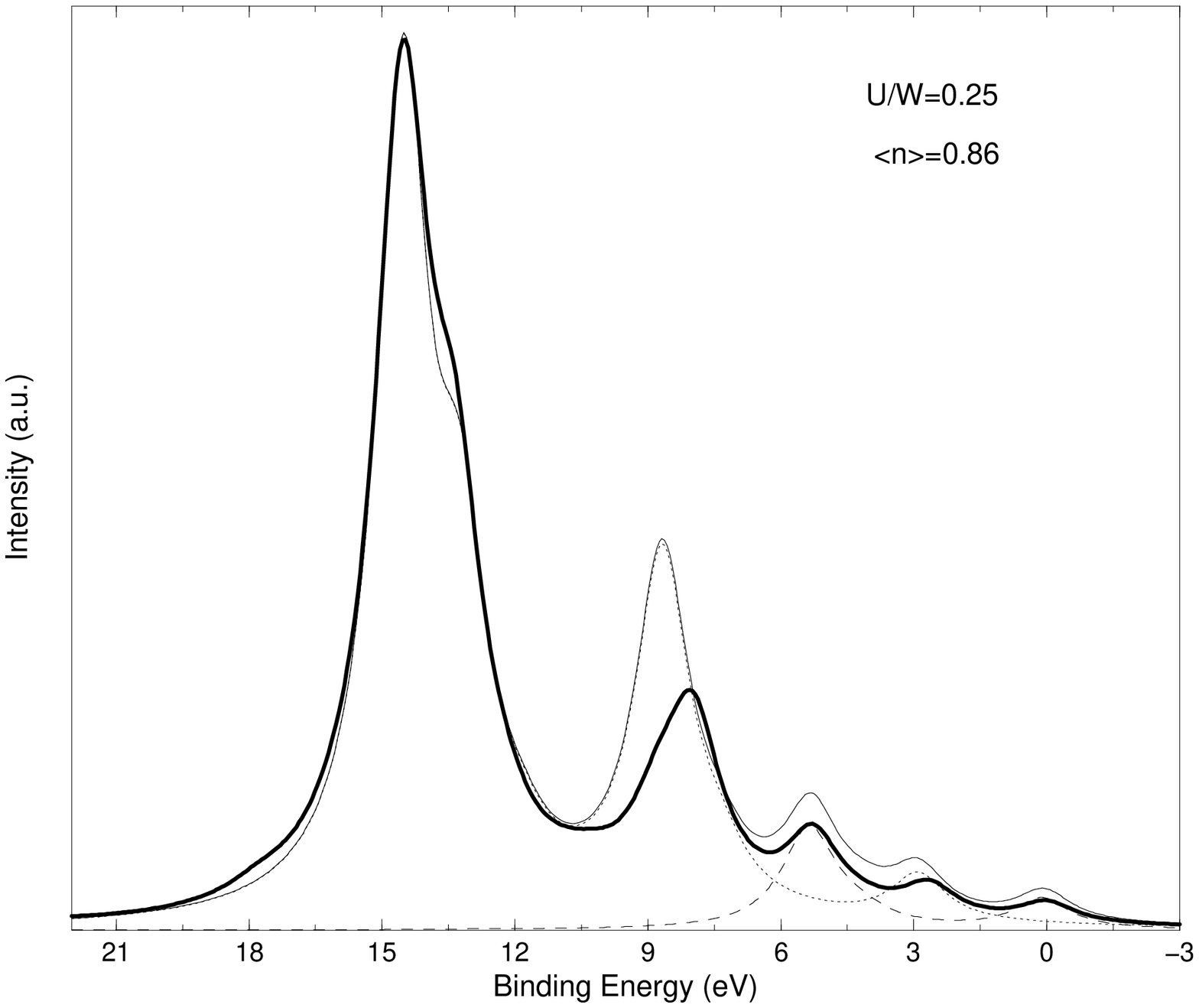,height=8cm,angle=-90}
\epsfig{figure=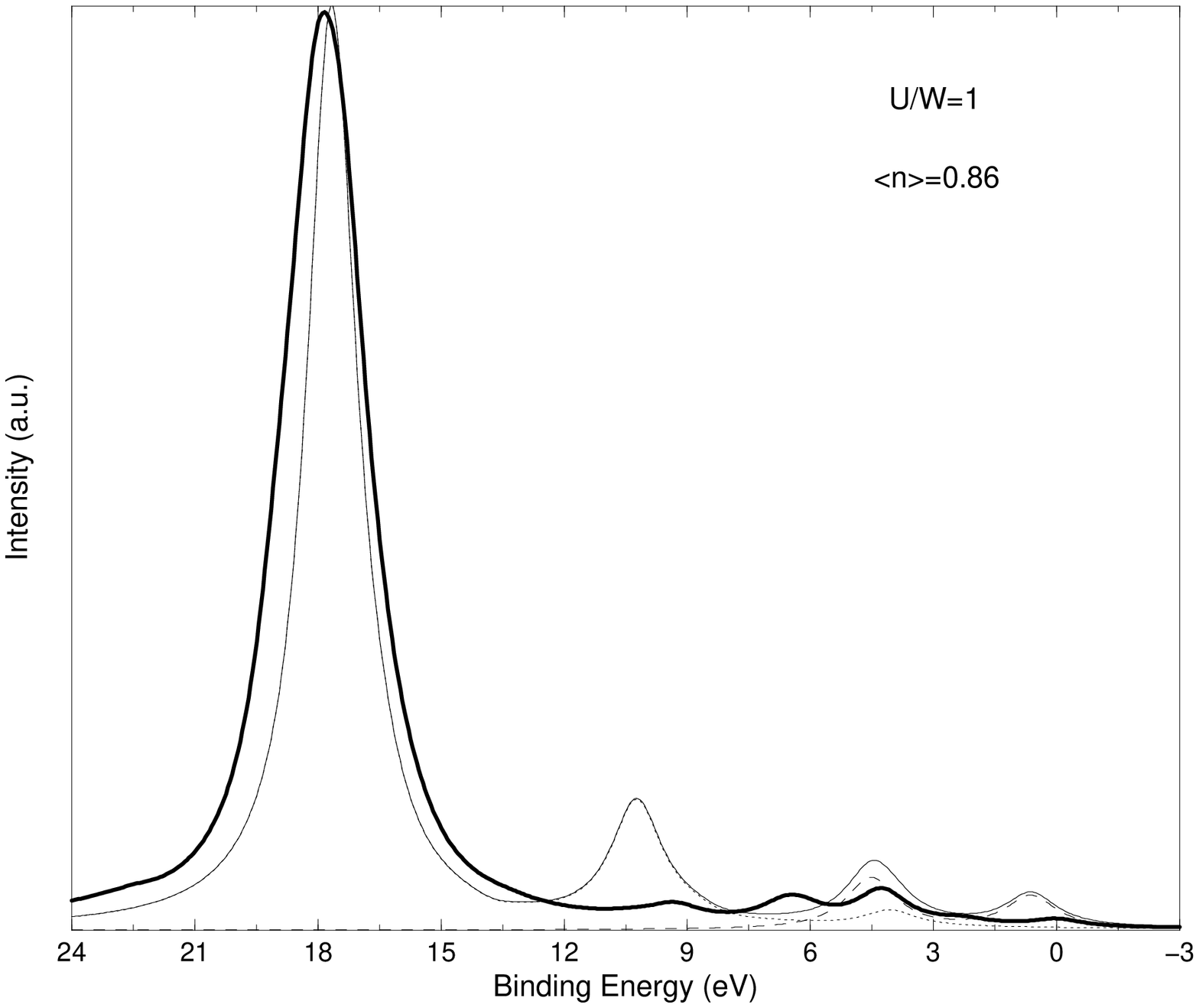,height=8cm,angle=-90}\\
\caption{\footnotesize{Binding energy dependence of $D_{2h1e}$ for 
$\langle n\rangle=0.86$ with  $U/W$=0.25 (left frame) and $U/W$=1 
(right frame). Heavy line: exact result; light: CLA;
dashed: 1-body contribution to the CLA result; dotted: 3-body
contribution to the CLA result. The line shapes have been 
convolved with a Lorentzian (FWHM=0.75 eV).}}\label{fig:7}
\end{center}\end{figure}

As one could expect, $D_{2h1e}(\go)$ has much more weight at high 
binding energies than $D_{1h}(\go)$, as one can see in Figure (\ref{fig:7}); 
in the right frame we have chosen the moderate coupling case $U/W$=0.25 with 
$\la n\ra=0.86$, like in Figure (\ref{fig:5}).

The overall agreement of the CLA with the exact results is quite good 
for the main peak, and even the small structures at lower binding energy are
reproduced in some detail. With increasing U, these structures are largely
lost (right frame of Figure (\ref{fig:7})); 
the CLA deteriorates in those regions where the intensity is low, 
but the overall line shape is still satisfactory.

The most remarkable feature is that the large increase of the 
interaction strenght produces a rather small increase of the binding 
energy of the dominant peak compared to the left frame  Figure (\ref{fig:7}). 
This would not be understandable if the peak were a two-hole resonance. 

Physically, the shift is refrained by the screening of the two holes by the
electron, which also becomes more effective with increasing $U$.  Our simple
approximation correctly accounts for the screening effect and consequently
predicts the position of the main peak quite accurately. It is not
surprising that $D_{2h1e}(\go)$ is more difficult to approximate in detail
than $D_{1h}(\go)$, since many more, highly  excited final states are 
reached with three quasiparticles than with one. 

\begin{figure}[t]
\begin{center}
\epsfig{figure=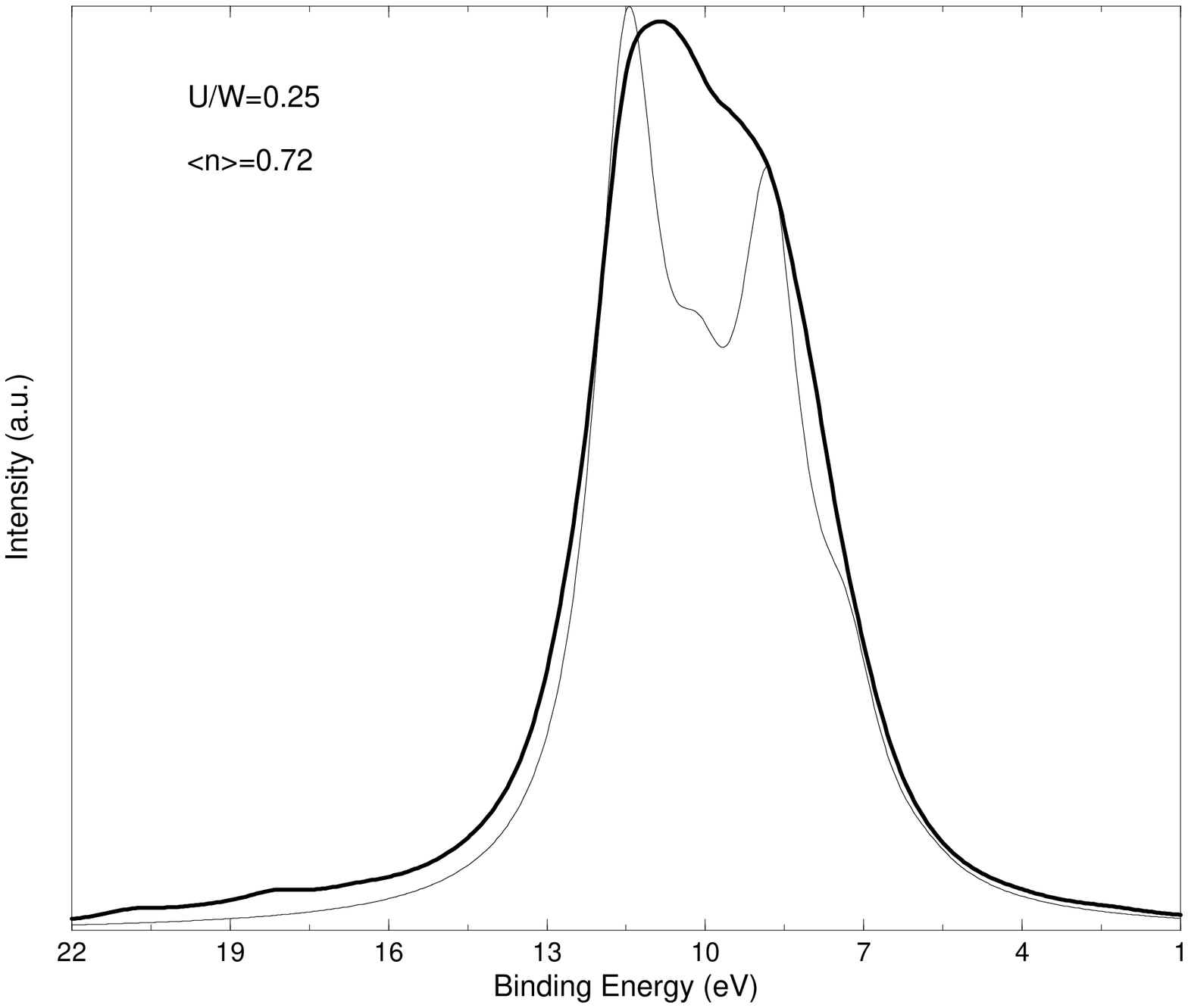,height=8cm,angle=-90}
\epsfig{figure=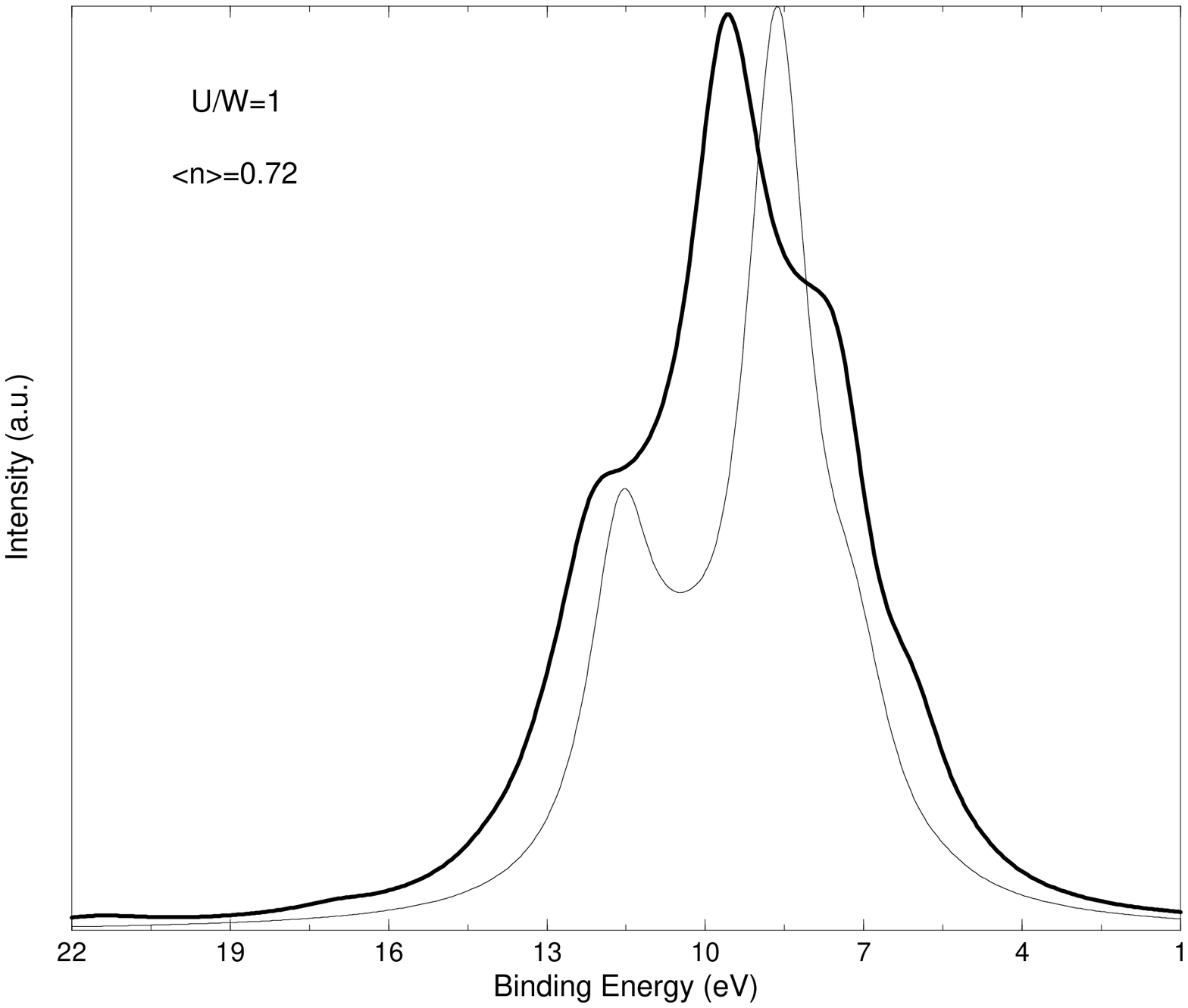,height=8cm,angle=-90}\\
\caption{\footnotesize{Binding energy dependence of $D_{2h1e}$ for
$\langle n\rangle=0.72$ with  $U/W$=0.25 (left frame) and $U/W$=1 (right frame).
Heavy line: exact result; light: CLA. 
The line shapes have been convolved with a Lorentzian (FWHM=0.75 eV).}}
\label{fig:8}
\end{center}\end{figure}

A particularly hard case is shown in Figure (\ref{fig:8}) where the filling 
is rather low ($\la n\ra=0.72$).
No single-particle contribuition exists in this case because
$\la\psi|a_{51\up}^{\dag}a_{52\up}|\psi\ra =0 $.

For $U/W$=0.25 (left frame) the agreement is fairly good.
For $U/W$=1 (right frame) the exact line shape shows three partially
resolved broad peaks covering a range of $\approx 10$ eV in binding
energy. The CLA in such severe conditions misses the main peak position
by $\approx 1$ eV and underevaluates the low binding energy shoulder;
however, we can still claim at least a qualitative agreement with the
results of the exact calculation. 

In all cases we find that the Herglotz property is fully preserved; 
this is a most valuable feature which is 
not easily obtained  for approximate three-body propagators. For 
instance, the approach of Ref.~\cite{Cini:94} fails in this respect 
at strong coupling.

Comparing the two frames of Figure (\ref{fig:8})  we observe
the apparent {\em negative-U} behaviour: increasing the interaction U,
the main peak shifts towards {\it lower} binding energies; this is  a
consequence of the interaction of the screening electron with the two Auger
holes.  We remark that a high enough $n_h$ is necessary to build up a 
localised screening cloud. This is why the {\em negative-U} behavior is 
observed in the early  transition metals, but not in the late ones.

\section{Conclusions}
To model Auger and APECS line shapes, we need a reliable practical 
recipe for a 3-body propagator, which poses much more difficult 
problems than the 1 and 2-particle Green's functions.
The Core-Ladder-Approximation that we are proposing produces a dramatic
simplification. The problem is reduced to the calculation of the mean-field 
one-body Green's function and to the diagonalisation of small matrices.
Comparison with exact model results shows 
that one can capture the essential  physics 
of a formidable problem which otherwise would require an excessive 
computational effort.

Experience with the 1 and 2-body problems suggests that the 3-body ladder 
series, without self-energy and vertex corrections, is a prosiming
approximation: it treats electrons and holes in the same way, and
produces non-negative densities of states.  
From a mathematical viewpoint, the CLA that we are proposing is the first step 
of a procedure which eventually leads to the exact summation of the
3-body ladder series. 
Like perturbation theory and other approaches to the many-body problem,
it allows systematic improvements, when required, at the cost of more 
computation. 
Physically, this simple approximation is well motivated and well balanced, 
and we provided evidence that it correctly describes the effect of the 
screening electron over the two final-state holes. 
The CLA is very accurate in the easy cases (high filling and/or 
small interaction); however, many transition metals are in the range of 
fillings and U where it is useful, and the approximation remains
reasonable, Herglotz and qualitatively good even for $U/W=1$. This holds
true for both the one-body features and the two-hole-one-electron
contributions which result from the full 3-body Green's function. The
property of the CLA of remaining qualitatively correct even at rather
strong coupling, when all other simple approaches badly fail, is an
appreciable feature. 
It depends on the fact that the Core Approximation becomes accurate at
strong coupling, which is the most critical {\em regime} for perturbation
theory; on the other hand, it allows to carry on the theory to all
orders, while treating the screening electron and the Auger holes on
equal footing. 
We are currently working out the application of the new approach to the
line shape analysis of actual experimental spectra.

\appendix
\section{The  Auger and APECS  currents}\label{appendix1}
\setcounter{equation}{0}
In a simplified form of the One-Step model~\cite{Gunn:1980} of the CVV 
Auger spectra, the current of electrons with energy $\ek$  measured in a 
Auger-Electron-Spectroscopy (AES) experiment is
\begin{align}
{\cal J}\(\ek\)=\int_{0}^{\infty}dt_{1}\int_{0}^{\infty}dt_{2} 
f(t_{1},t_{2}) e^{i \ek\(t_{1}-t_{2}\)}. \label{a1}
\end{align}
Here, letting $H_{S}$ represent the valence Hamiltonian without the 
core-hole, $| \psi\ra$ its ground state, $H'$ the valence Hamiltonian in 
the presence of the core-hole,
\begin{align}
f(t_{1},t_{2})=
\sum_{m,m^{\prime}} \la\psi|e^{i\(H'+i\Gamma\) 
t_{2}}|m\ra\la m|H_{A}^{\dag} e^{iH_{S}(t_{1}-t_{2})}H_{A}|m'\ra
 \la m'| e^{-i\(H'-i \Gamma\)t_{1}}|\psi\ra,
\label{a2}
\end{align}
where 
\begin{align}
H_{A}= \sum_{\ga,\gb}M_{\ga\gb}a_{\ga}a_{\gb}.
\label{a3}
\end{align}
produces the Auger holes with matrix elements $M$ in spin-orbitals 
denoted by Greek symbols; $\Gamma$ is an operator which produces 
virtual Auger transitions; core and free-electron operators have 
already been averaged out; the $m,m'$ summations run over a complete set 
of valence states. In Ref.~\cite{Cini:94} a simple approximation was 
proposed. The basic idea was that the complete set of summations are 
largely exhausted by summing over just two orthogonal states, namely, 
$|\psi\ra$ and  the relaxed initial state of the Auger transition,
$|\phi\ra$, that is, the ground state of the valence electrons in the
presence of the core-hole potential.
In this scheme, the Auger spectrum has two main contributions, {\em relaxed} 
and {\em unrelaxed}. The unrelaxed contribution arises from 
$m=m'=|\psi\ra$, and can be expressed in terms of the two-hole Green's 
function $G^{(2)}_{\go}$. The relaxed contribution arises from 
$m=m'=|\phi\ra$, and is proportional to
\begin{align}	
	\la\phi|a_{\ga}^{\dag}a^{\dag}_{\gb} e^{iH_{S}\(t_{1}-t_{2}\)}
	a_{\gb'}a_{\ga'}|\phi\ra.
	\label{eq:appA2}
\end{align}
By a variational calculation Cini and Drchal~\cite{Cini:94} showed that
\begin{align}
|\phi\ra \propto\sum_{\ga_{l}}a^{\dag}_{\ga_{l}}a_{F}|\psi\ra.
\label{eq:appA3}
\end{align}
with $a^{\dag}_{\ga_{l}}$ creates an electron in a localized spin-orbital at
the Auger site and $a_{F}$ annihilates an electron at the Fermi level. In
this way the screening cloud is represented by a single electron that 
has moved from the Fermi  surface to the empty local states of the Auger 
site.
Using Equations (\ref{eq:appA3}) and (\ref{eq:appA2}) one obtains an 
expression for the relaxed contribution to the Auger line shape, 
involving the 3-body Green's function of Equation (\ref{eq:g3}).

In APECS, on the other hand, one measures the Auger electron energy
distribuition in concidence with the photoelectron energy, and one 
can selectively study the decay of the relaxed hole state. 
Semi-empirical studies~\cite{future} show that for a broad range  of
photoelectron kinetic energies the summation over  $m,m'$ is dominated by
$|\phi\ra$.

\section{Proof of Equations (\ref{eq:core2}) and  (\ref{eq:core2}$'$)}
\label{appendix2}
\setcounter{equation}{0}
In first quantisation the operator $a^{\dag}_{\ga}a_{\gb}$ 
becomes $|\ga\ra\la\gb|$, and one can easily show that  
$a^{\dag}_{\ga}\(t\)a_{\gb}\(t'\)$ becomes 
$e^{iH_{0}t}|\ga\ra\la\gb|e^{-iH_{0}t'}$. The average of any one-body 
operator on a Slater determinant is the sum of the averages on the 
occupied spin-orbitals $k$. Therefore, letting $f_{k}=1$ for occupied 
states and  $f_{k}=0$ for empty ones,
\begin{align}
\la a^{\dag}_{\ga}\(t\)a_{\gb}\(t'\)\ra=\sum_{k}f_{k}e^{i\gee_{k}\(t-t'\)}
\la k|\ga\ra\la\gb|k\ra.\label{eq:a1}
\end{align}
Therefore, summing over all the atomic spin-orbitals, one gets
\begin{align}
\sum_{\gc}\la a^{\dag}_{\ga}\(t\)a_{\gc}\(t'\)\ra\la 
a^{\dag}_{\gc}\(t'\)a_{\gb}\ra=
\sum_{\gc}\sum_{k,k'}f_{k}f_{k'}e^{i\gee_{k}\(t-t'\)}
\la k|\ga\ra\la\gc|k\ra e^{i\gee_{k'}t'}
\la k'|\gc\ra\la\gb|k'\ra.\label{eq:a2}
\end{align}
Now, exploiting the completeness of the $\gc$ set, we obtain
\begin{align}
\sum_{\gc}\la a^{\dag}_{\ga}\(t\)a_{\gc}\(t'\)\ra\la 
a^{\dag}_{\gc}\(t'\)a_{\gb}\ra=
\sum_{k}f_{k}e^{i\gee_{k}t}\la k|\ga\ra
\la\gb|k'\ra=\la a^{\dag}_{\ga}\(t\)a_{\gb}\ra.\label{eq:a3}
\end{align}
In Equation (\ref{eq:a3}) the times $t,t'$ are arbitrary. We are free 
to insert $\theta$ functions to distinguish between positive and 
negative $t$ and ensure that $t'$ is intermediate between $t$ and 0.
This proves Equations (\ref{eq:core2}).

\section{Application of the CLA}\label{appendixc}
\setcounter{equation}{0}
From the diagonalisation of the one-body part of the problem, including the
Hartree-Fock terms (\ref{eq:cluster5}), one readily obtains the propagator 
$S^{h,e}_0\(i\gs,j\gs;t\)$ where $i=1,2$ and $\gs=\up,\dn$.
Since there are two {\em local} orbitals in this model and the matrix
$\left\{{\bf  T}_{ij}\right\}$ is real the set of equations 
(\ref{eq:coreans2}) represent a $3\times 3$ linear problem that determines
each component of the $R^{\pm}$ matrices, from which the  $B_{0}$ matrices
(\ref{eq:cla7a}) can be easly obtained.

For clarity, we report the system which determines  $R^+$: 
\begin{gather}
\(-i\)S^h_0\(1\gs,1\gs;t\)=R^+\(1\gs,1\gs,t\)S^h_0\(1\gs,1\gs;0^+\)+
R^+\(1\gs,2\gs,t\)S^h_0\(2\gs,1\gs;0^+\),\notag \\
\(-i\)S^h_0\(1\gs,2\gs;t\)=R^+\(1\gs,1\gs,t\)S^h_0\(1\gs,2\gs;0^+\)+
R^+\(1\gs,2\gs,t\)S^h_0\(2\gs,2\gs;0^+\),\\
\(-i\)S^h_0\(2\gs,2\gs;t\)=R^+\(2\gs,1\gs,t\)S^h_0\(1\gs,2\gs;0^+\)+
R^+\(2\gs,2\gs,t\)S^h_0\(2\gs,2\gs;0^+\),\notag
\end{gather}
yielding
\begin{align}
\begin{pmatrix}
R^+\(1\gs,1\gs,t\) \\ R^+\(1\gs,2\gs,t\) \\ R^+\(2\gs,2\gs,t\)
\end{pmatrix}
=\(-i\)\(
\begin{matrix}
\overline{n}_{11,\gs} & \overline{n}_{12,\gs} & 0 \\
\overline{n}_{12,\gs} & \overline{n}_{22,\gs} & 0 \\
0 & \overline{n}_{12,\gs} & \overline{n}_{22,\gs}
\end{matrix}\)^{-1}
\begin{pmatrix}
S^h_0\(1\gs,1\gs,t\) \\ S^h_0\(1\gs,2\gs,t\) \\ S^h_0\(2\gs,2\gs,t\)
\end{pmatrix}.
\label{eq:cluster10}
\end{align}
where $\overline{n}_{ij,\gs} \equiv S^h_0\(i\gs,j\gs;0^+\)$. The 
diagonal $\overline{n}$ elements are much bigger than the off-diagonal, 
and the problem is well posed.
A similar procedure gives the $R^-$ functions using the propagators 
$S^e_0\(i\gs,j\gs;t\)$.

We consider the following spin configurations:
\begin{align}
G\(ijk,k'j'i';\go\)\equiv
G\(i\up j\dn k\up,k'\up j'\dn i'\up;\go\).
\label{eq:cluster11}
\end{align}
where the $i,j,i^{\prime},j^{\prime}$ indices refer to the holes.
To find them, we write down explicitly the CLA equations 
(\ref{eq:cla7}), where only the interactions involving the orbitals 1 
and 2 at the Auger site $s=5$ appear. In the Hubbard model the interaction
terms are proportional to the product of number operators, and this entails
the condition
\begin{align}
U_{\mu\nu\gr\tau}=U\gd_{\mu,\gr}\gd_{\nu,\tau}.
\label{eq:cluster11a}
\end{align}
That simplifies the set of equations (\ref{eq:cla7}) 
that reduce to a $\(8\times 8\)$ linear system.
It is convenient to use the shorthand notation where 
underlined numbers mark the position of  $R^{\pm}$ functions, namely: 
\begin{gather}
G_0\(ij\underline{k},\underline{1}11;\go\)=
S^h_0\(i,1;\go\)S^h_0\(j,1;\go\)R^-\(k,1;\go\),\notag\\
G_0\(i\underline{j}k,1\underline{1}1;\go\)=
S^h_0\(i,1;\go\)R^+\(j,1;\go\)S^e_0\(k,1;\go\),\\
G_0\(\underline{i}jk,11\underline{1};\go\)=
R^+\(i,1;\go\)S^h_0\(j,1;\go\)S^e_0\(k,1;\go\),
\notag
\end{gather}
then, we may write the system in the form
\begin{multline}
G_0\(ijk,k'j'i;\go\)-G^{CLA}\(ijk,k'j'i;\go\)=\\
U\left\{\[G_0\(ij\underline{k},\underline{1}11;\go\)-
G_0\(\underline{i}jk,11\underline{1};\go\)\]
G\(111,k'j'i';\go\)+\capo
\[G_0\(ij\underline{k},\underline{2}11;\go\)-G_0\(i\underline{j}k,2
\underline{1}1;\go\)-
G_0\(\underline{i}jk,21\underline{1};\go\)\]G\(112,k'j'i';\go\)+ \capo
\[G_0\(ij\underline{k},\underline{1}21;\go\)-G_0\(\underline{i}jk,12
\underline{1};\go\)\]
G\(121,k'j'i';\go\)+ \capo
\[G_0\(ij\underline{k},\underline{2}21;\go\)-G_0\(i\underline{j}k,2
\underline{2}1;\go\)-
G_0\(\underline{i}jk,22\underline{1};\go\)\]G\(122,k'j'i';\go\)+ \capo
\[G_0\(ij\underline{k},\underline{1}12;\go\)-G_0\(i\underline{j}k,1
\underline{1}2;\go\)-
G_0\(\underline{i}jk,11\underline{2};\go\)\]G\(211,k'j'i';\go\)+ \capo
\[G_0\(ij\underline{k},\underline{2}12;\go\)-G_0\(\underline{i}jk,21
\underline{2};\go\)\]
G\(212,k'j'i';\go\)+ \capo
\[G_0\(ij\underline{k},\underline{1}22;\go\)-G_0\(i\underline{j}k,1
\underline{2}2;\go\)-
G_0\(\underline{i}jk,12\underline{2};\go\)\]G\(221,k'j'i';\go\)+ \capo
\[G_0\(ij\underline{k},\underline{2}22;\go\)-G_0\(\underline{i}jk,22
\underline{2};\go\)\]
G_0\(222,k'j'i';\go\)\right\}.
\label{eq:cluster12}
\end{multline}
For the $G^{sp}$ contribution  we use Equation (\ref{eq:picco3}) for 
the propagator $S\(j\dn,j'\dn;\go\)$,
\begin{align}
S\(j\dn,j'\dn;\go\)=S_0\(j\dn,j'\dn;\go\)-U^2\sum_{k\, k'}
S_0\(j\dn,k\dn;\go\)\Sigma\(k\dn,k'\dn;\go\)S\(k'\dn,j'\dn;\go\).
\label{eq:cluster13}
\end{align}
Finally, we obtain the proper self-energies $\Sigma\(k\dn,k'\dn;\go\)$  
from Equation (\ref{eq:picco3}), where we insert  the CLA form of the 
three body Green's function determined above. We recall that the one-body 
contributions arise from the anihilation of the up-spin electron and 
hole; therefore, one is left with the propagator for a down-spin 
hole, which is diagonal in spin because there are no spin-flip 
interactions in our model. Taking into account Equation 
(\ref{eq:cluster11a}), one finds  
\begin{multline}
\Sigma\(1\dn,1\dn;\go\)=G^{CLA}\(1\up 1\dn 1\up,1\up 1\dn 1\up;\go\)+
G^{CLA}\(2\up 1\dn 2\up,2\up 1\dn 2\up;\go\)+\\
2G^{CLA}\(1\up 1\dn 1\up,2\up 1\dn 2\up;\go\)+
G^{CLA}\(1\dn 2\dn 2\dn,2\dn 2\dn 1\dn;\go\),
\end{multline}
\begin{multline}
\Sigma\(1\dn,2\dn;\go\)=G^{CLA}\(1\up 1\dn 1\up,1\up 2\dn 1\up;\go\)+
G^{CLA}\(1\up 1\dn 1\up,2\up 2\dn 2\up;\go\)+\\
G^{CLA}\(2\up 1\dn 2\up,1\up 2\dn 1\up;\go\)+
G^{CLA}\(2\up 1\dn 2\up,2\up 2\dn 2\up;\go\)+\\
G^{CLA}\(1\dn 2\dn 2\dn,1\dn 1\dn 2\dn;\go\),
\end{multline}
\begin{multline}
\Sigma\(2\dn,2\dn;\go\)=G^{CLA}\(1\up 2\dn 1\up,1\up 2\dn 1\up;\go\)+
G^{CLA}\(2\up 2\dn 2\up,2\up 2\dn 2\up;\go\)+\\
2G^{CLA}\(1\up 2\dn 1\up,2\up 2\dn 2\up;\go\)+
G^{CLA}\(2\dn 1\dn 1\dn,1\dn 1\dn 2\dn;\go\),
\end{multline}
\begin{align}
\Sigma\(2\dn,1\dn;\go\)=\Sigma\(1\dn,2\dn;\go\).
\end{align}

\section*{Acknowledgments}
This work has been supported by the Is\-ti\-tu\-to Na\-zio\-na\-le di 
Fi\-si\-ca del\-la Ma\-te\-ria. One of us (A.M.) has been supported 
by the INFM scholarship ``A\-na\-li\-si te\-o\-ri\-ca del\-la Tec\-ni\-ca
A.P.E.C.S.'', prot.1350.




\begin{thebibliography}{99}

\bibitem{reviews} 	P. Weightman, Rep. Prog. Phys. {\bf 45}, 753 (1982);
			J. C. Riviere, Atomic Energy Reserch Report 
			AERE-R10384 (1982).
\bibitem{Cini:1976} 	M. Cini, Solid State Comm. {\bf 20}, 605 (1977).
\bibitem{Cini:1977} 	M. Cini, Solid State Comm. {\bf 24}, 681 (1977).
\bibitem{Sawatzky:1977}	G. A. Sawatzky, Phys. Rev. Lett. {\bf 39}, 504 (1977).
\bibitem{Au} 		C. Verdozzi, M. Cini, J. F. McGilp, G. Mondio, D. Norman, 
			J. A. Evans, A. D. Laine, P. S. Fowles, L. Du\`o and 
			P. Weightman, Phys. Rev. B {\bf 43}, 9550 (1991).
\bibitem{Ag} R.J.Cole, 	C. Verdozzi, M. Cini and P. Weigthmam, Phys. Rev. B 
			{\bf 49}, 13329 (1994).
\bibitem{offsite} 	C. Verdozzi, M. Cini, J. A. Evans, R. J. Cole, A. D. Laine,
			P. S. Fowles, L. Du\`o and  P. Weightman, Europhvsics 
			Letters {\bf 16} (8), 743 (1991);
			M. Cini and C. Verdozzi, Physica Scripta T {\bf 41}, 
			67 (1992);
			C. Verdozzi and M. Cini, Phys. Rev. B {\bf 51}, 
			7412 (1995).
\bibitem{cov} 		M. Cini, A. Pernaselci and E. Paparazzo, 
			J. Electron Spectroscopy and Rel. Phenomena {\bf 72}, 
			77 (1995);
			A. Pernaselci and M. Cini, J. Electron Spectroscopy and 
			Rel. Phenomena  {\bf 82}, 79 (1996).
\bibitem{Cini:1979} 	M. Cini, Surf. Sci. {\bf 87}, 483 (1979).
\bibitem{ds} 		S. Doniach and M. \v{S}unji\'{c}, J. Phys. C: Solid 
			State Phys. {\bf 3}, 285 (1970).
\bibitem{Galitzkii:1958}V. Galitzkii, Soviet Phys. JEPT {\bf 7}, 104 (1958).
\bibitem{Cini:89} 	M. Cini and C.Verdozzi,
			J. Phys. C: Condens. Matter {\bf 1}, 7457 (1989).
\bibitem{Cini:90} 	M. Cini, M. De Crescenzi, F. Patella, N. Motta, M. Sastry, F. Rochet,
			R. Pasquali, A. Balzarotti and C. Verdozzi, Phys. Rev. B {\bf 41}, 5685
			(1990).
\bibitem{Drchal:1984}  	V. Drchal and J. Kudrnovsky,
            		J. Phys. F: Metal Phys. {\bf 12}, 2443 (1984).
\bibitem{Verdozzi:86} 	M. Cini and C. Verdozzi, 
			Solid State Comm. {\bf 57}, 657 (1986).
\bibitem{graphite} 	J. E. Houston, J. W. Rogers, R. R. Rye, F. L. Hutson 
			and D. Ramaker, Phys. Rev. B {\bf 34}, 1215 (1986); 
			M. Cini and A. D'Andrea, in 
			{\em Auger Spectroscopv and electronic structure}, 
			Springer Series in Surface Science, Vol. 18 Edited by
			K. Wandelt, G. Mondio and G. Cubiotti 
			(Springer, Heidelberg, 1989).
\bibitem{expUneg}	P. Hedeg\aa rd and F.U. Hillebrecht, Phys. Rev. B
			{\bf 34}, 3045 (1986).
\bibitem{propUneg}	D. K. G. de Boer, C. Haas and G. A. Sawatzky, J. Phys. 
			F: Metal Phys. {\bf 14}, 2769 (1984).
\bibitem{Sarma:92}	D. D. Sarma, S. R. Barman, S. Nimkar and H. R. Krishnamurthy,
			Phys. Scri. T {\bf 41}, 184 (1992).
\bibitem{Gunn:1980}	O.Gunnarsson and K.Sch\"{o}nhammer,
			Phys. Rev. B {\bf 22}, 3710 (1980).
\bibitem{Cini:94}	M. Cini and V. Drchal,	
			J. Phys. C: Condens. Matter {\bf 6}, 8549 (1994);
			M. Cini and V. Drchal, J. Electron Spectroscopy and 
			Rel. Phenomena {\bf 72}, 151 (1995).
\bibitem{Haak:1978} 	H. W. Haak, G. A. Sawatzky and T. D. Thomas, 
		    	Phys. Rev. Lett. {\bf 41}, 1825 (1978).
\bibitem{Haak:1984}	H. W. Haak, G. A. Sawatzky, L. Ungier, J. K. Gimziewsky and 
			T. D. Thomas, Rev. Sci. Instrum. {\bf 55}, 696 (1984).
\bibitem{Sawatzky:book}	G. A. Sawatzky, 
			{\it Auger photoelectron coincidence spettroscopy},
			Treatise on material science and technology, Vol. 30
		  	Accademic Press. New York.
\bibitem{Jensen}	E. Jensen, R. A. Bartynski, S. L. Hulbert and E. D. Johnson,
			Rev. Sci. Instrum. {\bf 63}, 3013 (1992).	
\bibitem{Gunnarsson:82} O. Gunnarsson and K. Sch\"{o}nhammer,
			Phys. Rev. B {\bf 26}, 2765 (1982).
\bibitem{sl} 		G. A. Sawatzky and A. Lenselink, Phys. Rev. B {\bf 21}, 
			1390 (1980).
\bibitem{Mattuck}  	R. D. Mattuck, {\em A guide to Feynman diagrams in the
			Many-Body problem}, McGraw-Hill, New York (1976) 
			Chapter 10.
\bibitem{future}	M. Cini and A. Marini, to be published.
		
\end{thebibliography}
\end{document}